\documentclass[apj]{emulateapj}
\usepackage[T1]{fontenc}
\usepackage{ae,aecompl}

\newcommand{\ie}{{\it i.e.,~}}
\newcommand{\eg}{{\it e.g.,~}}

\begin{document}
%===============================================================================
\title {Why Do Only Some Galaxy Clusters Have Cool Cores?}
%===============================================================================
\author{Jack~O.~Burns\altaffilmark{1}, Eric~J.~Hallman\altaffilmark{1,2}, Brennan Gantner\altaffilmark{1},Patrick~M.~Motl\altaffilmark{3},Michael~L.~Norman\altaffilmark{4}}
\altaffiltext{1}{Center for Astrophysics and Space Astronomy, Department of Astrophysical \& Planetary Science, University of Colorado, Boulder, CO 80309}
\altaffiltext{2}{National Science Foundation Astronomy and
  Astrophysics Postdoctoral Fellow}
\altaffiltext{3}{Department of Physics and Astronomy,
Louisiana State University, Baton Rouge, LA 70803} 
\altaffiltext{4}{Center for Astrophysics and Space Sciences, University of California-San Diego, 9500 Gilman Drive, La Jolla, CA 92093}
\email{jack.burns@cu.edu}

\begin{abstract}
Flux-limited X-ray samples indicate that about half of rich galaxy
clusters have cool cores.  Why do only some clusters have cool cores
while others do not?  In this paper, cosmological N-body + Eulerian
hydrodynamic simulations, including radiative cooling and heating, are
used to address this question as we examine the formation and
evolution of cool core (CC) and non-cool core (NCC) clusters.  These
adaptive mesh refinement simulations produce both CC and NCC clusters
in the same volume.  They have a peak resolution of 15.6 $h^{-1}$ kpc
within a $(256~ h^{-1} Mpc)^3$ box.  Our simulations suggest that
there are important evolutionary differences between CC clusters and
their NCC counterparts.  Many of the numerical CC clusters accreted
mass more slowly over time and grew enhanced cool cores via
hierarchical mergers; when late major mergers occurred, the CC's
survived the collisions.  By contrast, NCC clusters experienced major
mergers early in their evolution that destroyed embryonic cool cores
and produced conditions that prevented CC re-formation.  As a result,
our simulations predict observationally testable distinctions in the
properties of CC and NCC beyond the core regions in clusters. In
particular, we find differences between CC versus NCC clusters in the
shapes of X-ray surface brightness profiles, between the temperatures
and  hardness ratios beyond the cores, between the distribution of
masses, and between their supercluster environs.  It also appears that
CC clusters are no closer to hydrostatic equilibrium than NCC
clusters, an issue important for precision cosmology measurements. 

\end{abstract}
%===============================================================================
%===============================================================================
\section{Introduction}
Clusters of galaxies with ``cool cores'' have peaked X-ray emission
(\ie excess above that extrapolated inward from a $\beta$-model fit of
the X-ray profile beyond the core) coincident with supergiant
elliptical galaxies.  They have central cooling times typically ${<0.1
{H_o}^{-1}}$ and central gas temperatures $\approx$30-40\% of the
virial temperatures \cite[\eg][]{Ikebe97,lsb02,pete03}.  Cores of
cool gas are found to be common in flux-limited samples, although a
selection bias is likely present because of the strongly peaked X-ray
emission in these clusters.  From a sample composed of clusters detected with {\it Einstein}, \citet{white97} found cool cores (CC) in $\approx$60\% of
their 207 cluster sample.  From a sample of 55 flux-limited {\it
ROSAT}-observed clusters, \citet{peres} found CC's in over
70\% of their galaxy clusters.  More recently, \citet{chen07}
identified 49\% of their 106 clusters as having cool cores in a
flux-limited sample, HIFLUGCS, based upon both {\it ROSAT} and {\it ASCA}
observations. 

Why do some, but not all, galaxy clusters contain cool cores?  To answer this question, we must explore the origin and evolution of cool cores galaxy clusters.  The earliest and simplest model assumed clusters to be spherical, isolated systems where ``cooling flows'' formed; as radiating gas loses
pressure support, cooling gas flows inwards to higher density values
which further accelerates the cooling rate \cite[\eg][]{fabian02}.  But,
the predicted end-products of this mass infall (\eg star formation,
HI, CO) have not been observed and the central temperatures indicate
that the gas at the cores has only moderately cooled \cite[see review by][]
{dv04}.  The current paradigm calls for heating to offset cooling,
possibly by AGNs via strong shocks \cite[\eg][]{heinz98} or weak
shocks \cite[\eg][]{fabian03}, or by AGNs $+$ thermal conduction
\cite[\eg][]{rusz04}, or by AGNs $+$ preheating
\cite[\eg][]{mccarthyheat}, or by Alfv{$\acute{e}$}n waves induced by
AGNs in the inner core and cluster mergers in the outer cores
\cite[\eg][]{fujita07}. 
  
The simple cooling flow model did not incorporate the important
effects of mergers and on-going mass accretion from the supercluster
environment in which these clusters reside
\cite[\eg][]{motl04,poole06}.  \citet{burns97} and \citet{gomez02}
first examined the result of cluster collisions in 2-D numerical
simulations that involved two idealized spherical clusters with
$\beta$-model density profiles and central cool cores that collided
together head-on.  They found that the ram pressure from major mergers
(\ie subcluster to cluster mass ratios of $\approx$15\% to 100\%) tends to disrupt the cool cores.
Similarly, \citet{rt02} and \citet{rick01} found
disruptions of cool cores by major mergers between spherical clusters
using 3-D simulations.  These numerical models may suggest that the
numbers of cool cores diminish as clusters grow via mergers (\ie
fewer cool cores in richer clusters at smaller $z$). 

More recently, we performed numerical simulations of the formation and
evolution of clusters in a cosmological context using the adaptive
mesh refinement N-body/hydro code {\it Enzo}, aimed at further
understanding cool cores \citep{burns04,motl04}.  The gas
in these clusters was evolved with radiative cooling but no heating.
We found that cooling modifies not only the cores but also
significantly alters the cluster appearance out to the virial radius
\cite[see also][]{akah}.  As new subcluster halos fall into a
cluster, they gradually donate cool gas so that the cool cores grow
over time.  Most mergers are oblique with halos spiraling into the
cluster centers and gently bequeathing cool gas to enhance the cores.
Thus, in this model, cool cores themselves grow hierarchically via the
merger/accretion process.  This model predicts that even cool core
clusters should possess a variety of substructures such as bullet subclusters and cold fronts, similar to those
observed in Abell clusters \cite[\eg][] {hallmark,bullet, shocks}.  It
also suggests cool cores may grow stronger (\ie cooler and denser) as
rich clusters increase in mass at recent epochs. 

The \citet{motl04} simulations were limited by the baryonic physics
that included only radiative cooling.  This model suffers from the
well-known cooling catastrophe \cite[\eg][]{whiterees} that results
in an overproduction of cool cores and an increase in the baryon
fraction \cite[\eg][]{knv05}.  Nearly every dark
matter potential well in this simulation was occupied by a halo of gas
that had cooled significantly.  Furthermore, these cool cores are
``hard'', generally denser, colder, and with more distinct boundaries than are observed.  The steep density contrast shelters the cores from ram pressure stripping, thus allowing them to survive and grow robustly during mergers with other halos.  Clearly, a more realistic model of cool cores must involve
added physical processes that ``soften'' the cores, thus making some
susceptible to disruption during mergers.  Heating by star formation \cite[\eg][]{valdarnini} or by AGNs would potentially soften the cores.  Additional softening effects may include thermal conduction \cite[\eg][]{zaknar,rusz02} along with mixing/heating from radio jet/lobe entrainment, and weak shocks and turbulent heating arising from halo mergers
\cite[\eg][]{burns98,bk02,fujita,vd05,math06}.  

In this paper, we present {\it Enzo} cosmology simulations that include radiative cooling, star formation (\ie a mass
sink for cold gas), and heating \cite[see also][]{motl05,hall06}.
Unlike  previous simulations, our cooling $+$ heating prescription has
succeeded in producing both cool core and non-cool core clusters
within the same computational volume.  In addition to the somewhat
more realistic baryonic physics, these simulations have the advantage
of bigger volumes and larger samples of clusters than in previous
computational simulations \citep{motl04, knv05}.  Thus, we have the
dataset to examine evolutionary effects in these numerical clusters
and can address the question in the title of this paper with good
statistics.  In Section 2, we describe the new simulations and the
analysis of the numerical clusters.  In Section 3, we compare the
statistical properties of our numerical clusters with recent observed
samples and show that the agreement is good.  In Section 4, we
describe new insights into the formation of cool core (CC) and
non-cool core (NCC) clusters from our simulations.  In Section 5, we
describe the observational consequences of evolutionary differences in
CC and NCC clusters.  Conclusions and a summary are presented in
Section 6.  

\section{Numerical Simulations}

The simulations described in this paper were performed with the {\it Enzo}  \footnote{http://lca.ucsd.edu/portal/software/enzo} code \citep{oshea04}
that couples an N-body algorithm for evolving the collisionless dark
matter particles with an Eulerian hydrodynamics scheme (PPM) that
utilizes adaptive mesh refinement.  We adopt a ${\Lambda}$CDM
cosmology with ${\Omega}_b$ = 0.026, ${\Omega}_m$ = 0.3,
${\Omega}_{\Lambda}$ = 0.7, h = 0.7, $n_s$=1, and ${\sigma}_8$ =
0.9. The simulation was initialized at z=30 using the CDM transfer
function from \cite{powerspectrum}. A low-resolution simulation was
first used to identify clusters in a volume of 256 $h^{-1}$ Mpc on a
side using 128$^{3}$ dark matter particles and grid zones.
High-resolution simulations were then performed that evolved the
entire volume but adaptively refined 50 smaller regions (as separate
simulations) around the largest clusters identified on the low
resolution grid. Each of these 50 regions first is refined with two
levels of static nested grids, each having a cell size half that of
its parent grid (thus each region has spatial resolution 4 times
better than the parent grid).  Within the nested grids, the dark
matter particles have a mass resolution of $9 \times 10^9 h^{-1}
M_{\odot}$. Then, within the nested static grids, we evolve the
simulation with 5 additional levels of adaptive refinement, again with
a factor of two increase in spatial resolution at each level. Cells
are flagged for refinement based on the local baryonic and dark matter
overdensities, refining on thresholds of 8.0 times the minimum value
at that level.  The spatial resolution on the finest grid is $15.6
h^{-1}$ kpc, adequate to resolve the cool core ($r_{CC} \approx 100
h^{-1}$ kpc), but not to probe the details of its structure \cite[see
  also][]{motl04}.  

Radiative cooling is calculated from a tabulated cooling curve derived
from a Raymond-Smith plasma emission model \citep{brickhouse95}
assuming a constant metallicity of 0.3 relative to solar.  The cooling
curve is truncated below a temperature of $10^4$ K.  Every timestep,
we calculate the energy radiated from each cell and remove that amount
of energy from the cluster gas \citep{motl04}. 

As mentioned above, star formation provides one mechanism to soften
cool cores by both transforming rapidly cooling gas into star
particles (and, therefore, removing the cold gas) and by heating the
surrounding gas with energy injected from supernovae.  The star
formation and heating that we used follows the prescription outlined
by \citet{cen92} and described in \citet{burns04}.  In brief, the code
examines all grid cells at the finest refinement level above a
specified overdensity.  The gas is converted to collisionless ``star''
particles when it is undergoing compression, rapid cooling, and the
mass in the cell exceeds the Jean's mass.  The star formation rate
is coupled to the local dynamical time and to a user-specified star
formation efficiency.  Once formed, the new star particle deposits
energy in the gas to simulate the instantaneous feedback from Type II
supernovae.  The strength of the supernova feedback is controlled by
another efficiency parameter, $\epsilon$, which gives the thermal
energy injected in proportion to the estimated rate of star formation
for that particle ($\dot{M}_{star} = M_{star} / t_{dyn}$, and $\dot{e}
= \epsilon \dot{M}_{star} c^2$).  The most important parameter in the
star formation recipe was found to be the strength of thermal feedback
from prompt supernovae.   Through trial and error, we found that the
value of $\epsilon = 4.11 \times 10^{-6}$ yields a reasonable fraction
of baryons locked in star particles at the current epoch \cite[see
  also][]{burns04}.  This value for the feedback parameter corresponds
to (for a star formation rate of one solar mass per year) a supernovae
rate of one per century with an average energy generation of $7 \times
10^{50}$ ergs per supernova. The chosen star formation parameters also
produce both CC and NCC clusters in the same volume. 

\begin{figure}
\begin{center}
%\epsscale{0.5}
\includegraphics[width=2.4in]{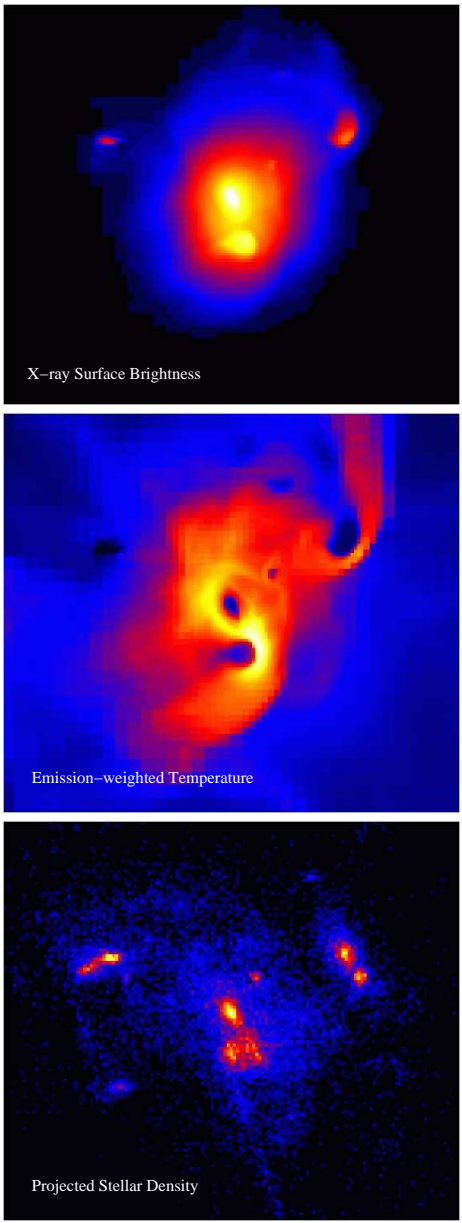}

\caption{Representative example of a cool core cluster with $M_{200} = 5 \times 10^{14} M_{\odot}$ and $T_{virial} =$ 4.2 keV at $z=0$ from the AMR simulation volume.  The top image is projected synthetic X-ray surface brightness.  The middle image is emission-weighted temperature (blue is $T < 4$ keV and yellow is $T > 5$ keV).  The bottom image is a projection of the star particle density. The field of view is 3.5 $h^{-1}$ Mpc.}
\label{CC_3panel}
\vspace{-10mm}
\end{center}  
\end{figure}

In Figure \ref{CC_3panel}, we show a representative example of a cool
core cluster at $z=0$ including images of the bolometric X-ray surface
brightness, emission-weighted temperature, and the distribution of
star particles (see \cite{hall06} for details on construction of
synthetic X-ray and temperature images).  These images illustrate the
dynamic range in X-ray structures and temperatures typical in
simulations of CC clusters, including the off-center infall of lower
mass cool halos with leading bow shocks.  The star particle image
shows the distribution of sinks of cold gas and extended heating as
new halos are accreted. 

The average total energy injection rate for the 10 most massive CC and
10 NCC clusters with a comparable mass distribution (at $z=0$) within
our computational volume is $\approx 5 \times 10^{43}$ ergs/sec. This
is comparable to the X-ray luminosity for these clusters, thus our
prescription produces an approximate balance between heating and
cooling.  This energy injection rate is also similar to the typical
kinetic luminosities thought to power radio jets/lobes in central
cluster radio sources \cite[\eg][]{burns90, eilek06, gentile07,
  wise07}.  Thus, we view this energy injection scheme as a
generalization of pre-heating of the core from a variety of sources,
including AGN.  The average energy injection rate is approximately the
same over all epochs between $z \approx 1$ and $z=0$. This feedback is
also comparable for CC and NCC clusters of the same mass suggesting,
as we show in Section 4, that something other than feedback, namely
mergers, drive the evolution of these two cluster types. 

We constructed a catalog of all numerical clusters with $M_{200} >
10^{14} M_{\odot}$ from z=0 to z=2.  ($M_{200}$ is measured out to the
radius, $r_{200}$, where the density is 200 times the critical density
and is $\approx M_{virial}$ which we will use interchangeably
throughout this paper.)  At $z=0$, we have 94 clusters in the sample,
but the entire catalog out to $z=2$ contains 1522 clusters (many are
the same cluster but at different epochs) giving us one of the largest
samples of numerical rich clusters to date from a cosmological
simulation.  These cluster simulations are publicly archived
\footnote{http://lca.ucsd.edu/data/sca}.  We have constructed a master
table of the basic properties of these clusters including the average
emission-weighted temperatures, virial and gas masses, $r_{200}$
($\approx$ the virial radius), baryon fractions, $\beta$-model fit
parameters, CC or NCC designation, and other properties.  The archive
and this table will be presented in \citet{hall07}.  These archived
clusters form the basis of the analysis of numerical clusters
presented in this paper. 

After visually inspecting all the temperature profiles for the $z=0$
clusters, we defined a cool core cluster to be one that has a
$\ge$20\% reduction in the central temperature compared to the
surrounding region (where the slope of the temperature profile becomes
negative; see Figure \ref{transition}) and this candidate baryonic
cool core is within one zone ($\approx$16 $h^{-1}$ kpc) of the dark
matter density peak.  This is a conservative definition that will
yield the smallest number of cool cores, but we estimate that more
liberal definitions will not increase the number by more than $\approx
10$\%.  With this strict classification, we find that 16\% of all the
numerical clusters with $M_{200} > 10^{14} M_{\odot}$ have cool cores.
This is low relative to the most recently observed fraction of 49\%
\citep{chen07} that comes from a flux-limited sample (which may be
biased somewhat high by flux boosting from the cool cores).  Several
possible effects may be operating to reduce the fraction of numerical
CC clusters.  First, our baryon fraction (2.6\%) for this simulation
is now recognized as low relative to the recent value from WMAP III
(4.2\%, \cite{spergel07}).  A higher gas fraction could result in more
robust cool cores.  Second, the power spectrum normalization
($\sigma_8$) may play a role in determining the number of CC clusters
(our current value of $\sigma_8$ is larger than that inferred from
WMAP III).  Third, the numbers of cool cores and their survival during
mergers appear to be a sensitive function of the heating/cooling
prescription.  Fourth, numerical resolution is likely a factor in the
production of cool cores.   

\section{Statistics of the X-ray Properties of CC and NCC Numerical Clusters}
How well do the general characteristics of our numerical clusters
match real galaxy clusters?  This is an important question to address
before we propose a new formation scenario for cool core and non-cool
core clusters based upon our numerical simulations. 

In the analysis that follows, we calculated the projected average temperatures for our simulated clusters as ``spectroscopic-like" temperatures as in \citet{rasia}. The weighting of temperature is different from the standard
emission-weighted temperature, and has been shown to be more consistent with the value of the temperature which would be deduced from an X-ray spectral fit. The calculation performs 
\begin{equation}
T_{500 SL} = \frac{\int n^2 T^a/T^{1/2} dV}{\int n^2 T^a/T^{3/2} dV}, 
\end{equation}
where a=0.75 is the value fitted from \citet{mazz04} which best
approximates the value of the spectroscopic temperature from X-ray
fitting. In our case we have integrated this weighting in a cylinder
with a radius of $r_{500}$ around the cluster center.  

\begin{figure}
\begin{center}
\includegraphics[width=3.5in]{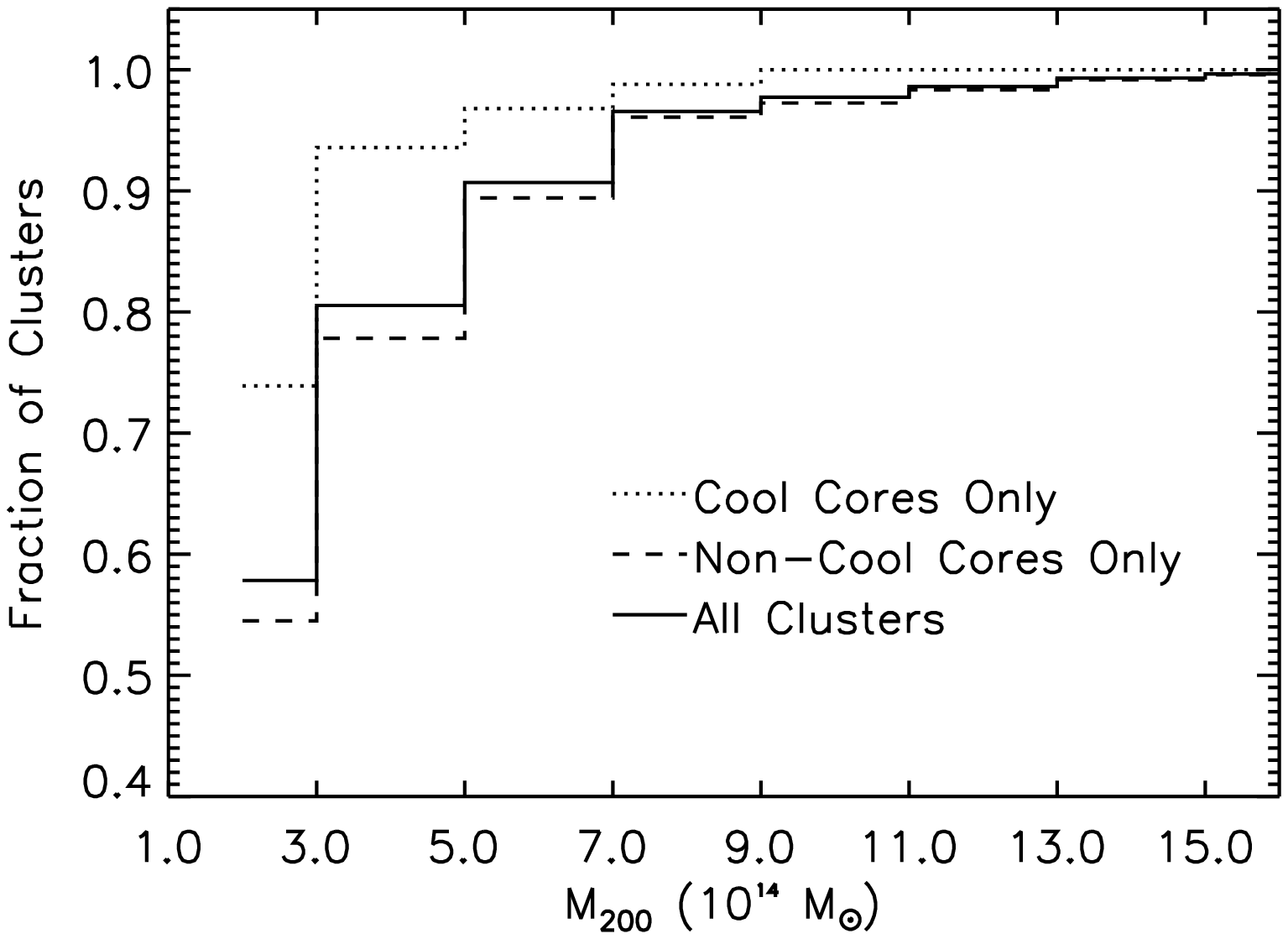}
\includegraphics[width=3.5in]{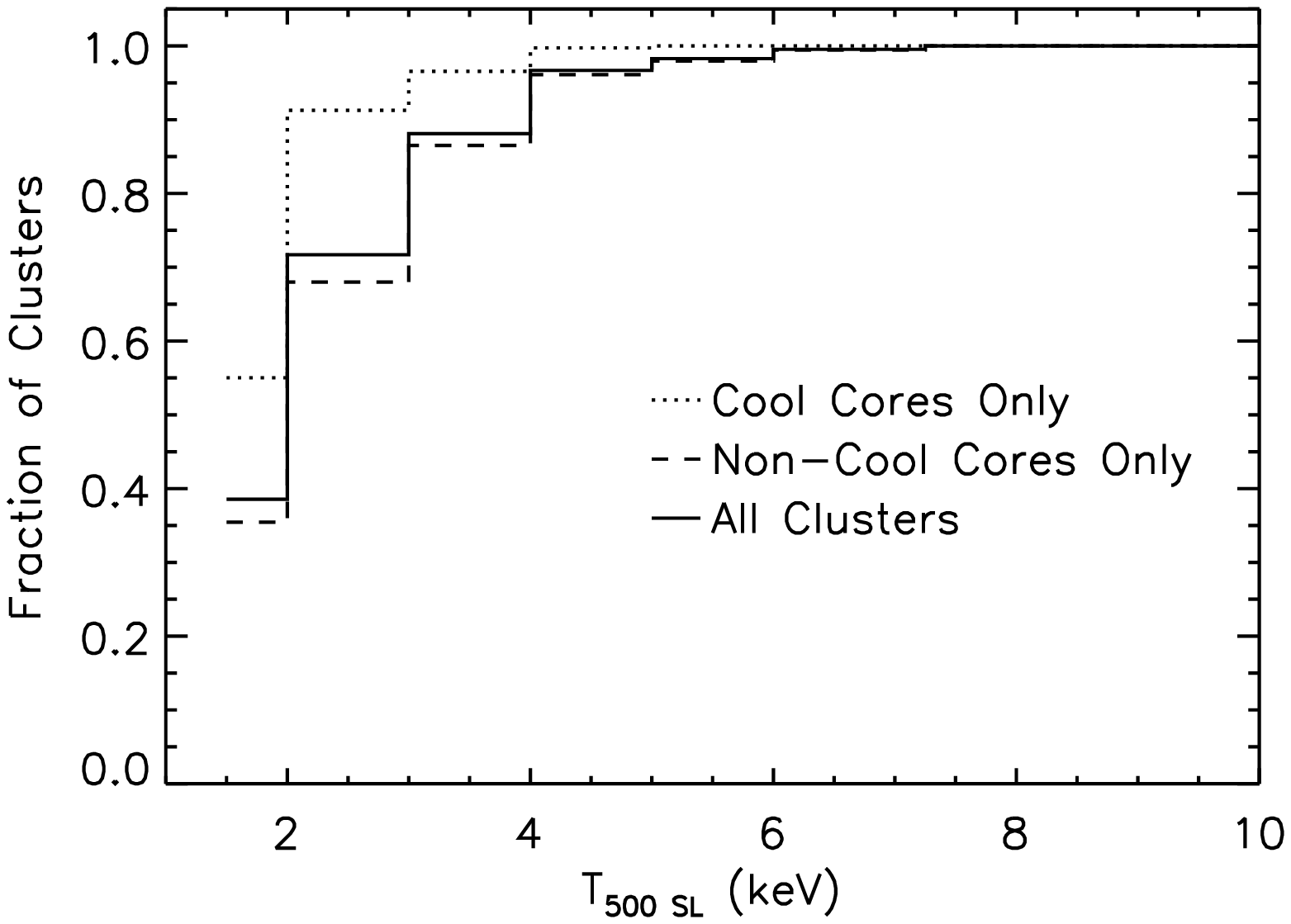}

\caption{Distribution of the cumulative fractions of total cluster masses and $T_{500 SL}$ (spectroscopic-like temperatures within $r_{500}$) for all clusters (solid) and for each cluster type in the catalog with $M_{200} > 10^{14} M_{\odot}$ and $z < 1$ (to approximately match present range of X-ray cluster   observations).} 

\label{masstemp}
\vspace{-8mm}
\end{center}  
\end{figure}

In Figure \ref{masstemp}, the distributions of $M_{200}$  and $T_{500 SL}$ (spectroscopic-like temperature inside $r_{500}$) for the
overall catalog of clusters, as well as for CC and NCC clusters within
the catalog, are shown.  We eliminated the cool core regions in
calculating $T_{500 SL}$ so as not to bias these temperature measurements and to use a technique similar to that applied to observations.  As will be discussed further in Section 5, there are fewer high mass, high temperature CC clusters in comparison to NCC clusters in the catalog. At $z=0$, the average mass of the CC clusters is $2.4 \pm 1.4 \times 10^{14} M_{\odot}$ and that of the NCC clusters
is $4.7 \pm 3.4 \times 10^{14} M_{\odot}$. 

\begin{figure}
\begin{center}
\includegraphics[width=3.5in]{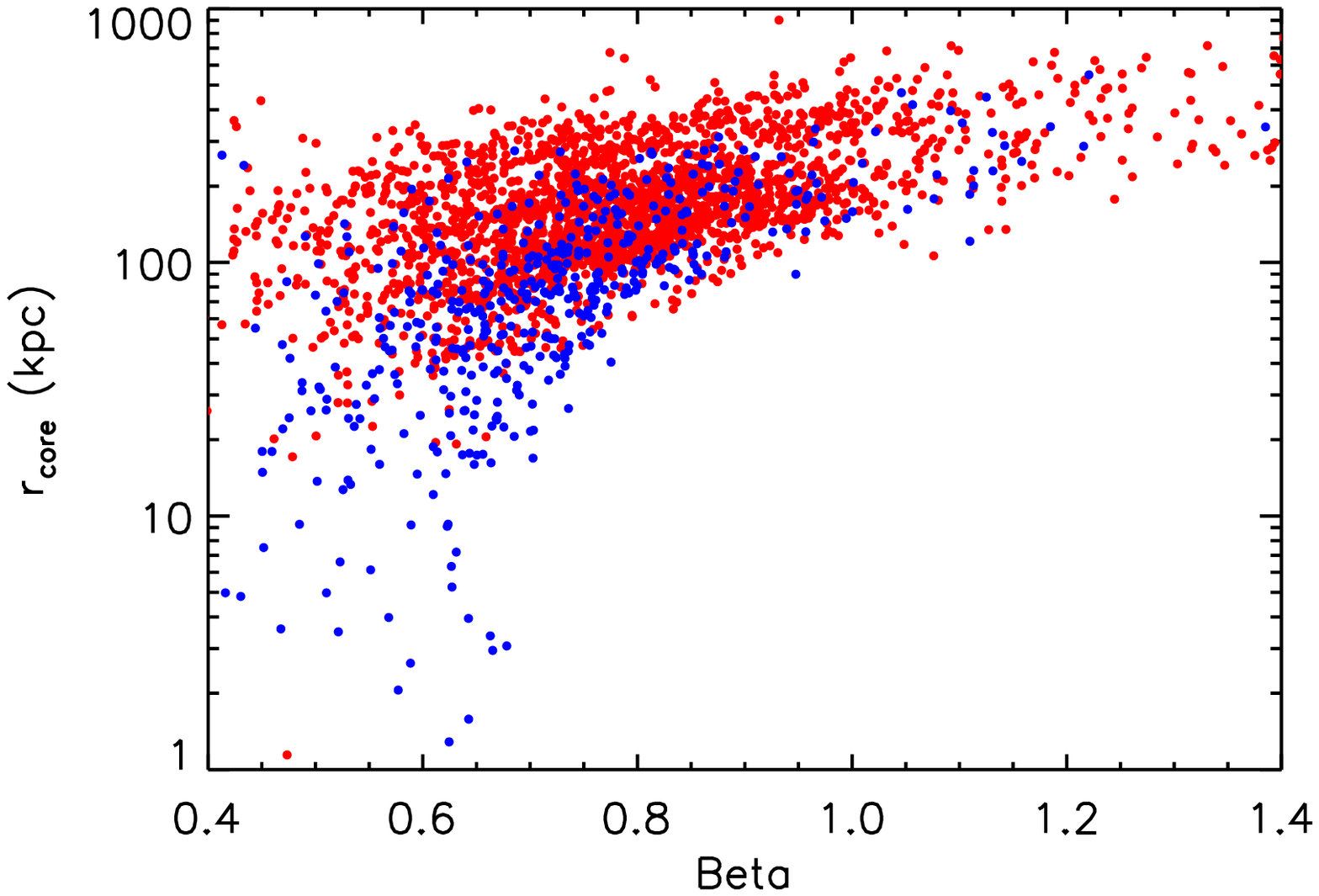}
\includegraphics[width=3.5in]{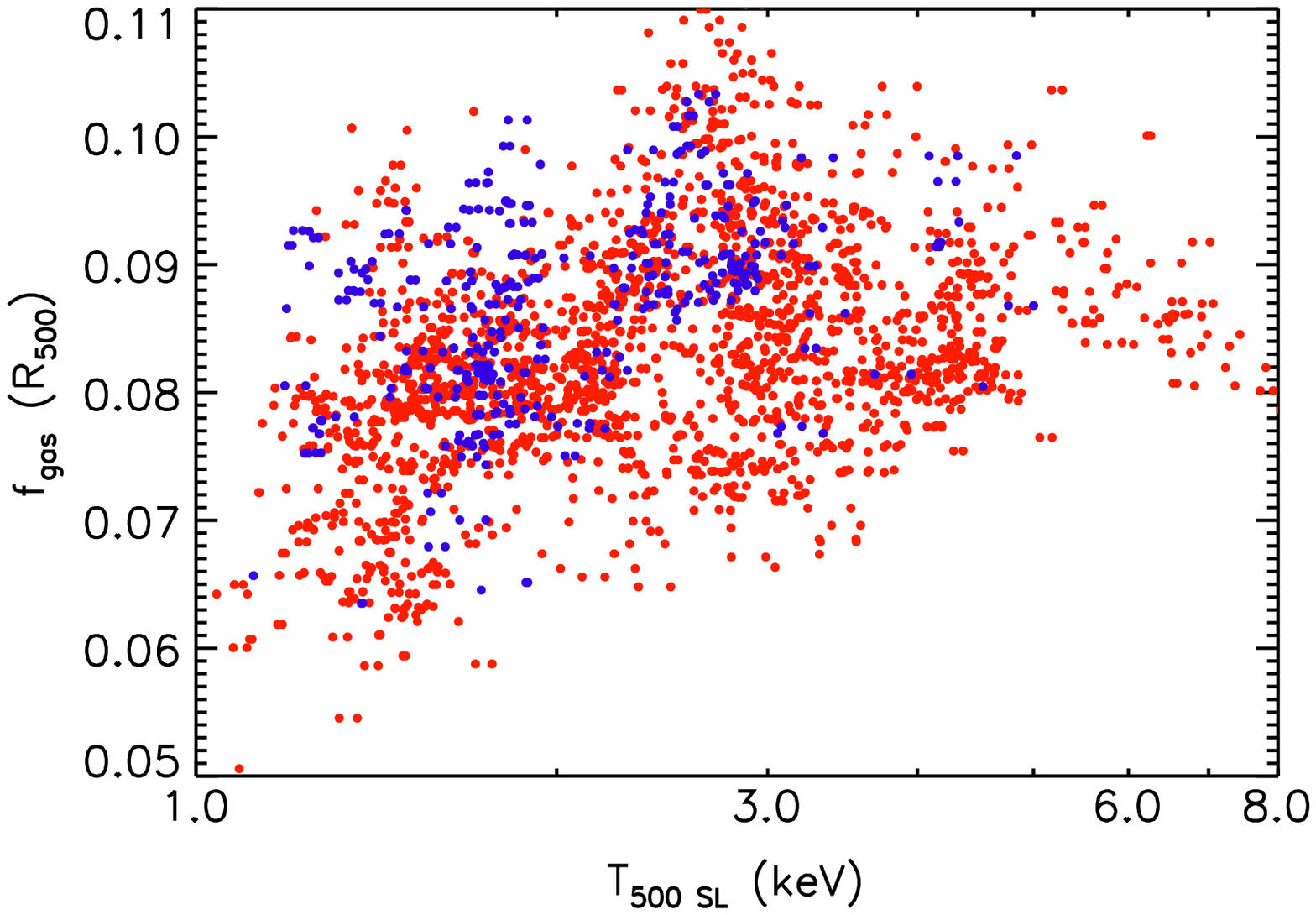}
\includegraphics[width=3.5in]{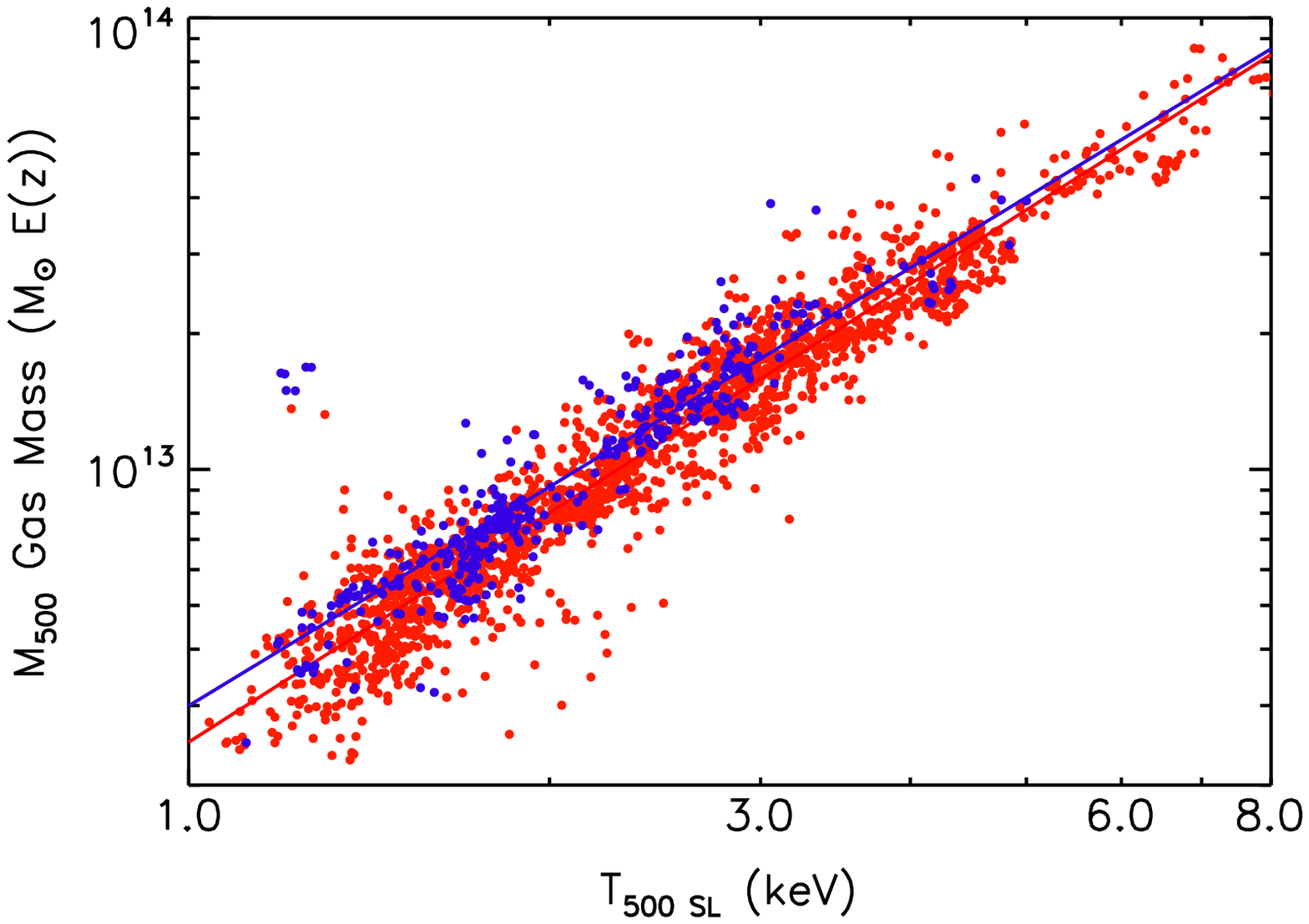}

\caption{Statistical properties of numerical galaxy clusters with $M_{200} > 10^{14} M_{\odot}$ and $z < 1$. Blue are CC and red are NCC clusters.  Top: $\beta$-model fits to $S_X$ profiles of individual clusters provide a measure of the core radius ($r_{core}$) and slope ($\beta$).  Middle: Gas fraction versus   $T_{500 SL}$ (spectroscopic-like temperatures within $r_{500}$).  Bottom: Gas mass out to $r_{500}$ is plotted against $T_{500 SL}$.  Power-law fits, performed separately for CC and NCC clusters, are shown.} 

\label{statistics}
\vspace{-8mm}
\end{center}  
\end{figure}

In Figure \ref{statistics}, we show examples of the most reliable
statistics and basic relationships between variables that are
typically calculated from X-ray observations.  We compare these
numerical data for $z<1$ clusters (to approximately match the range of
redshifts for current observations) in our catalog with those obtained
from the recent statistically complete sample of clusters observed
with {\it ROSAT} and {\it ASCA} as reported by \citet{chen07}.  The
numerical clusters were separated into CC and NCC using the criteria
noted in Section 2.  In the top panel of Figure \ref{statistics}, we
plot the core radius versus the slope ($\beta$) for a $\beta$-model
fit to the synthetic X-ray surface brightness profile ($S_X \propto
[1+(r/r_c)^2]^{1/2 -3\beta}$). For the CC clusters, the cool cores
were excluded from the fit (see Section 5.2 for details).  This plot
shows a separation between CC and NCC clusters such that cool core
clusters have smaller cores, $r_c$, for a given $\beta$.  This
separation and the overall results from these $\beta$-model fits match
up very well with Figure 3 in \citet{chen07}. 

The second panel shows the mean gas fraction measured out to $r_{500}$
($f_{gas}(r_{500}) = \langle \rho_{gas}/\rho_{total} \rangle_{500}$)
as a function of $T_{500 SL}$.  We attempted to correct our gas fractions by multiplying $f_{gas}$ by the ratio of $\Omega_b$ measured by WMAP III to that which we used in these simulations (Section 2). This brings our gas fractions into somewhat better agreement but we emphasize that they are still low relative to recent observations \cite[\eg][]{mccarthy07, relaxed06, sadat05}.  We do find good qualitative agreement in the shape and distribution of points in this figure relative to Figure 13 in \citet{chen07}.  There may be a slight tendency for reduced gas fractions at lower temperatures, as in observations
\cite[\eg][]{lin03, mccarthy07}, but $f_{gas}$ is otherwise constant
for $T > 3$ keV.  There is a hint of a weak separation between CC and
NCC clusters with cool cores having somewhat higher gas fractions for
a given temperature as we will discuss in Section 4. 

The third panel presents a plot of gas mass out to $r_{500}$ against
$T_{500 SL}$.  There is a strong scaling relation with comparable
power-law slopes for each type of cluster (measured slope index of
$1.61 \pm 0.04$ for CC clusters and $1.69 \pm 0.01$ for NCC clusters).
This scaling relation is qualitatively similar to that in Figure 11
from \citet{chen07}, although the slope is a bit steeper than that
observed and expected for self-similar behavior \citep{kaiser86} (\ie
$M \propto T^{1.5}$).  \citet{relaxed06} find a flatter slope for the
$M$\textendash $T$ relation for their sample of 13 cool core clusters
in comparison to other authors who analyzed mixed samples with CC and
NCC clusters (and different techniques for measuring mass and
temperature).  

In a separate paper \citep{jeltema07}, we also show that the
distribution of X-ray substructure within these clusters, as measured
using power ratios, agrees with that observed from X-ray observations
of nearby rich galaxy clusters.  

Overall, within the noted limitations of these simulations, the
average properties and the relationships between basic variables for
the numerical clusters agree fairly well with X-ray observations.

\section{The Formation of CC and NCC Clusters}

With a relatively large sample of numerical clusters, we are able to
explore the question of why some clusters have cool cores but others
do not.  These simulations indicate that the evolution of CC clusters
has followed a different history in terms of accretion of mass from
the cosmic web in comparison to NCC clusters.  Although each cluster
has its own unique rich and complex evolutionary path that depends
upon its initial mass and the density of surrounding halos, there are
some clear general trends within this larger dispersion that we see which appear to separate CC from NCC clusters. 

\begin{figure}
\begin{center}
\includegraphics[width=3.5in]{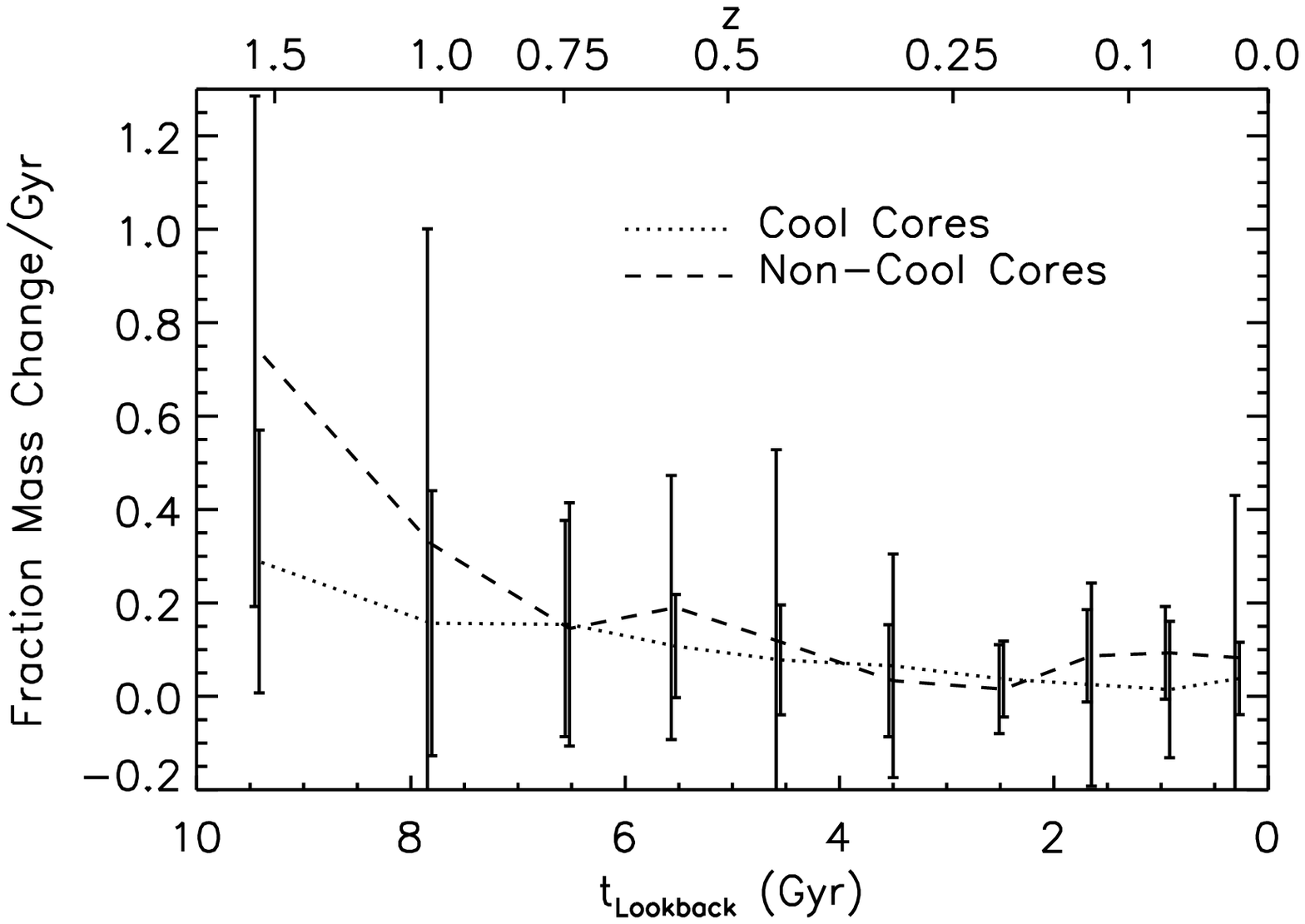}
\includegraphics[width=3.5in]{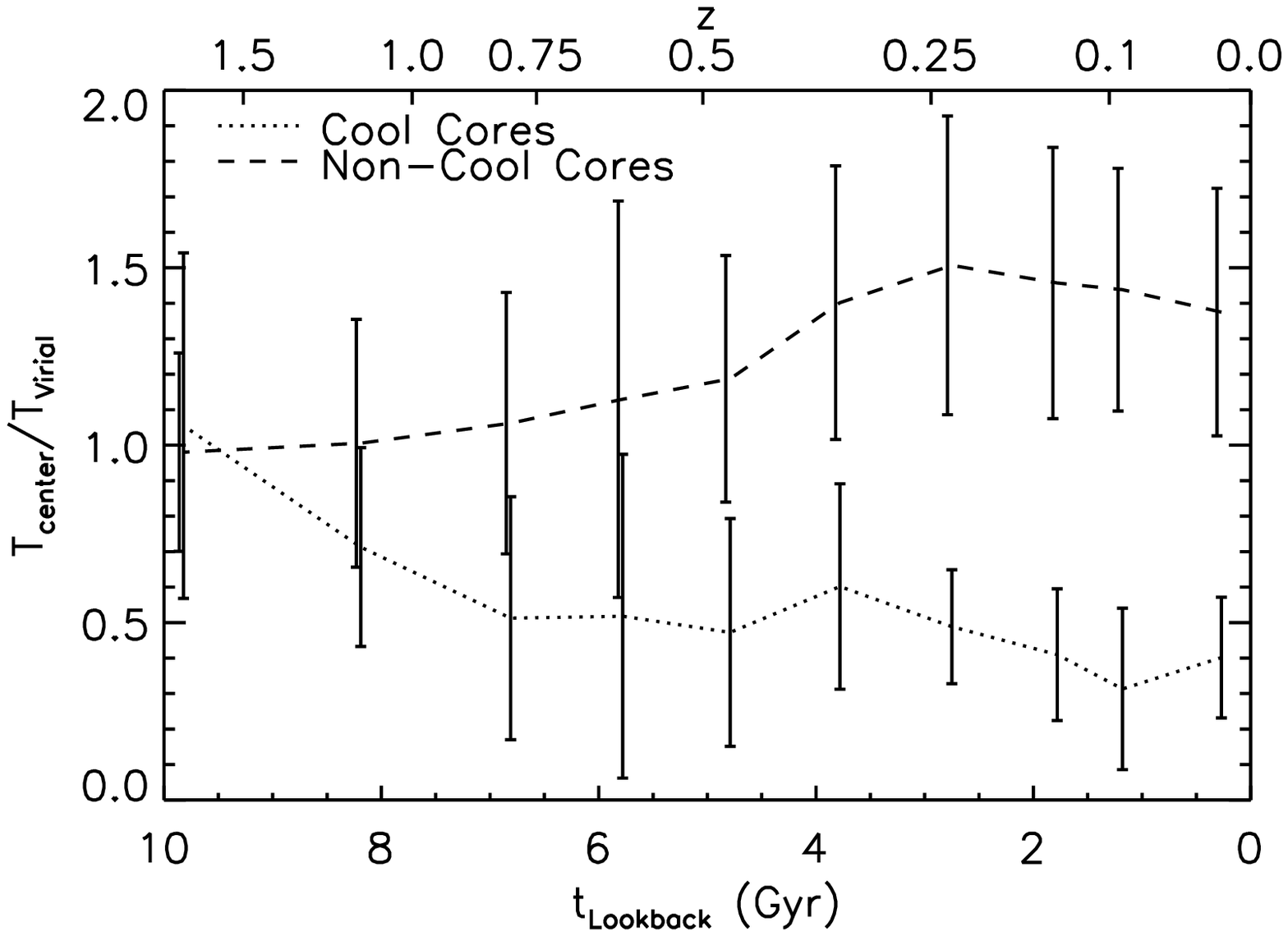}

\caption{The median time evolution of the 10 most massive NCC
  clusters and the 10 most massive CC clusters. Dispersions in the distributions are also shown.  Top: The median fractional
  mass change per Gyr as a function of lookback time and redshift.
  Bottom: Evolution of the median core temperatures (normalized by the virial
  temperature).  The NCC clusters accrete significantly more mass than
  the CC clusters until $z \approx 0.5$, signifying more early major
  mergers than for the CC clusters.  By $z=0.25$, the core
  temperatures for the NCC clusters are about 3 times hotter than the
  CC clusters; the cool cores are well established and becoming more
  robust (slightly cooler and denser) throughout subsequent minor
  mergers.} 
\label{evol}
\vspace{-10mm}
\end{center}  
\end{figure}

In Figure \ref{evol}, we track the evolution of the median changes in
cluster mass ($\dot{M}_{200}$) and the median central temperatures for the 10
most massive CC and NCC clusters from $z=0$ back to between $z=1$ and
$z=2$ (each cluster is tracked back only to a time determined by our
mass cutoff of $M_{200} > 10^{14} M_{\odot}$).  Although we use the ten highest masses in each sample for better statistics, we note that the same
qualitative trends as shown in Figure \ref{evol}, but with larger dispersion, are present for samples of CC and NCC clusters selected to have comparable mass
distributions.  We illustrate this evolution back to just $z \approx 1.5$ because there are only a few CC clusters at $z>1.5$ above our $10^{14} M_{\odot}$ mass limit (more clusters grow above this mass over time).

The two cluster types show different histories in their median mass accretion rates at early times. At $z=1.5$, NCC clusters experience a median $\approx$75\% change in mass per Gyr, albeit with a not unexpected large dispersion due to the wide range of merger states.  The CC clusters have a median change of $\approx$30\% in mass per Gyr with a smaller dispersion but also for fewer clusters with $M_{200} > 10^{14} M_{\odot}$. Using a K-S test, we find that for $1<z<1.5$, the distributions of mass change for NCC and CC clusters differ at the 95\% level.

{\it The NCC clusters demonstrate a trend of experiencing major mergers early   in their histories up to $z \approx 0.5$, which destroy any initial cool cores, then they settle down to a more quiescent state thereafter}.  Here we define a ``major'' merger as one that has the potential for disrupting a nascent cool core, usually accreting $\ge$50\% of the cluster's previous mass over a timescale of $\approx$1 Gyr.  {\it CC clusters, on the other hand, avoid mergers   with high fractional mass changes early in their histories and instead grow slowly such that the cool cores increase in mass and stability.}  As shown in Figure \ref{evol}, CC clusters after $z \approx 0.5$ have a relatively constant rate of accretion continuing to the present, similar to NCC clusters.

The central temperature plot
in Figure \ref{evol} demonstrates that similar starting conditions can
result in either CC or NCC clusters.  At early epochs, the dispersion in central temperatures is large and the distributions are statistically indistinguishable between what will become CC and NCC clusters at $z=0$.  This contrasts to the significant difference in central temperatures between CC and NCC clusters for $z < 0.5$.  Thus, the early merger history
primarily determines the eventual cluster configuration at the present
epoch. 

\begin{figure*}
\begin{center}
%\epsscale{0.5}
%\includegraphics[width=2.5in]{f5a.eps}
%\includegraphics[width=2.5in]{f5b.eps}
%\includegraphics[width=2.5in]{f5c.eps}
%\includegraphics[width=2.5in]{f5d.eps}
\includegraphics[width=5.0in]{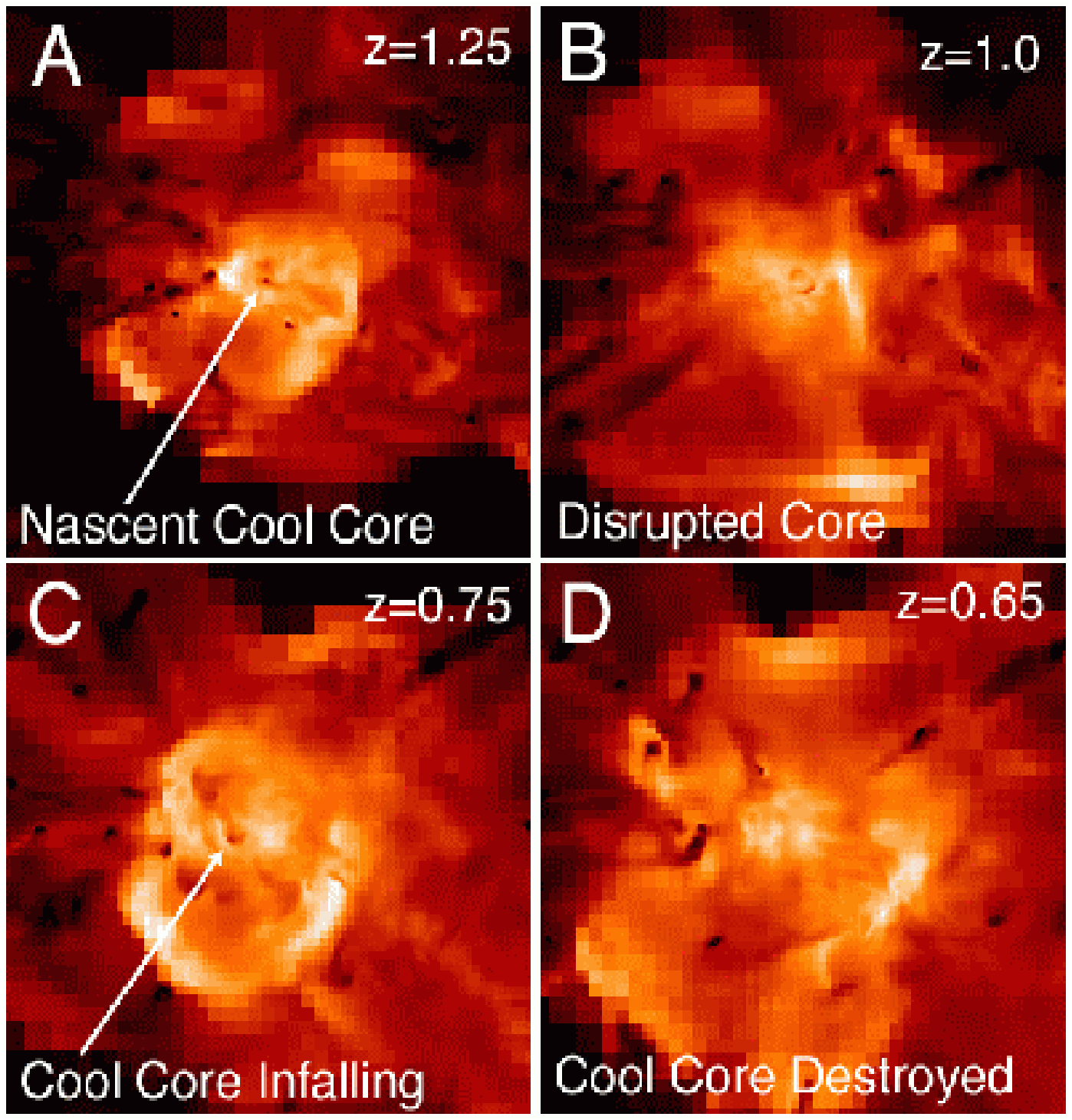}
\includegraphics[width=3.5in]{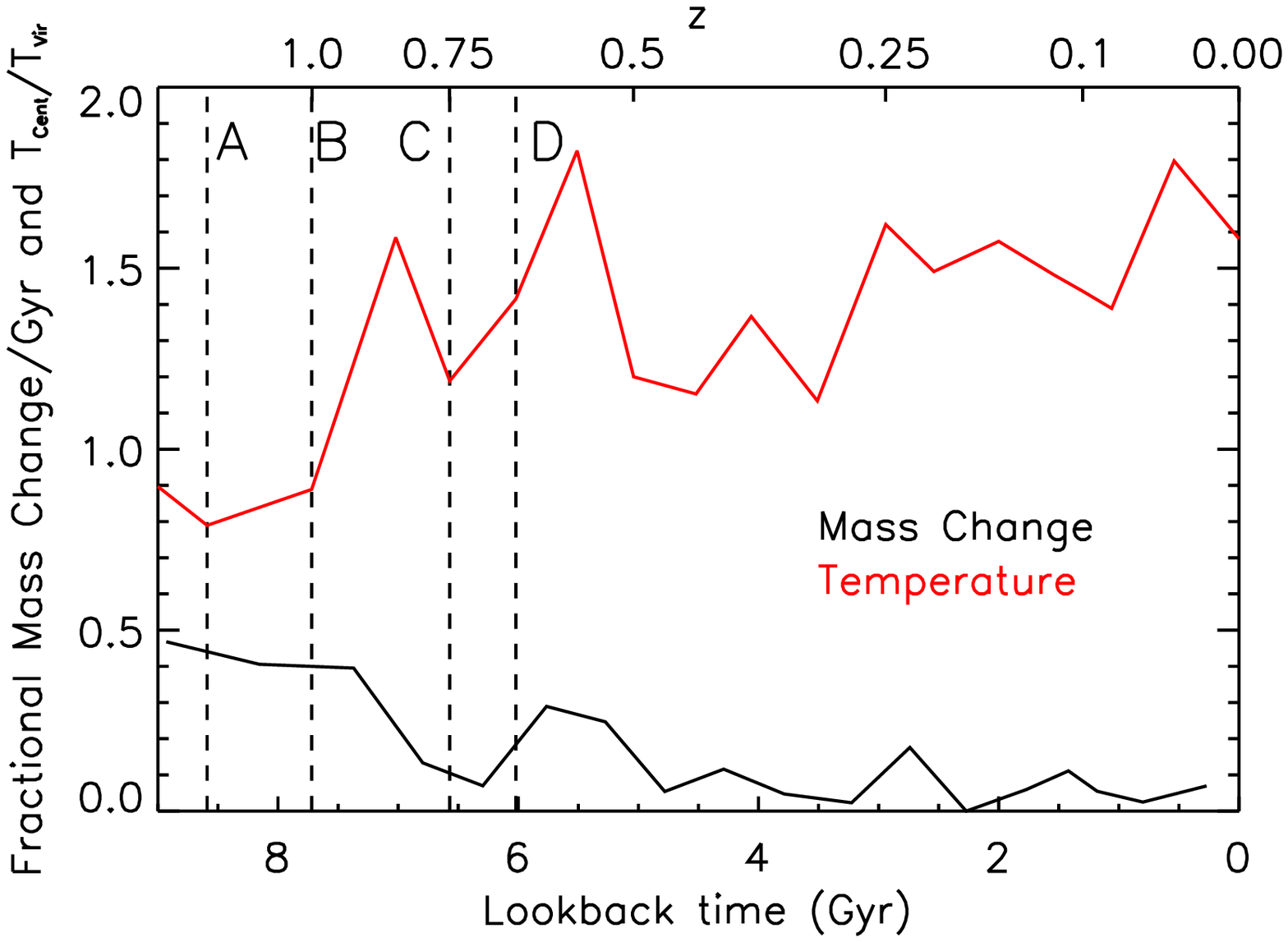}

\caption{Four snapshots of the history of an NCC cluster with final
  ($z=0$) values of $M_{200} = 8 \times 10^{14} M_{\odot}$ and
  $T_{virial} =$ 5.5 keV.  Dashed vertical lines in the bottom panel
  correspond to different epochs of the temperature images.} 
\label{NCCevol}
\vspace{-8mm}
\end{center}  
\end{figure*}

\begin{figure*}

\begin{center}
\includegraphics[width=5.0in]{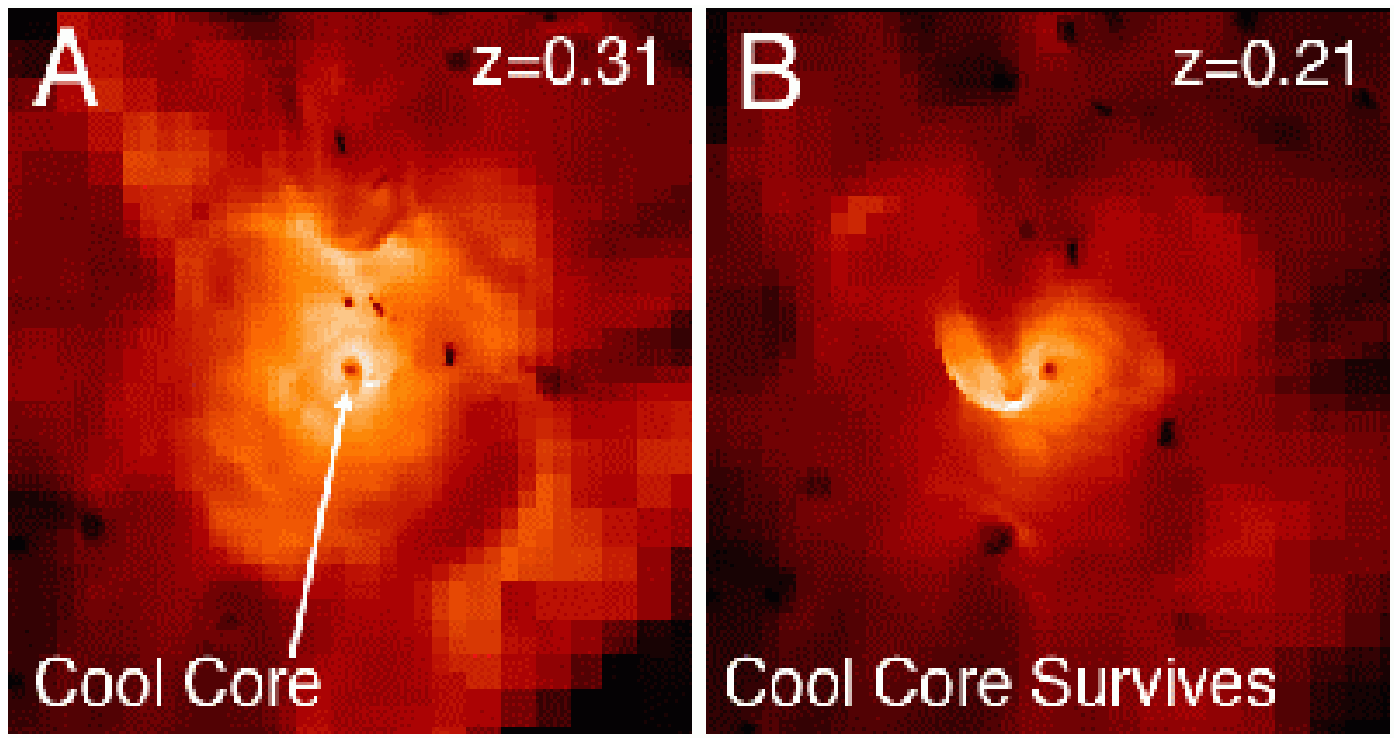}
\includegraphics[width=3.5in]{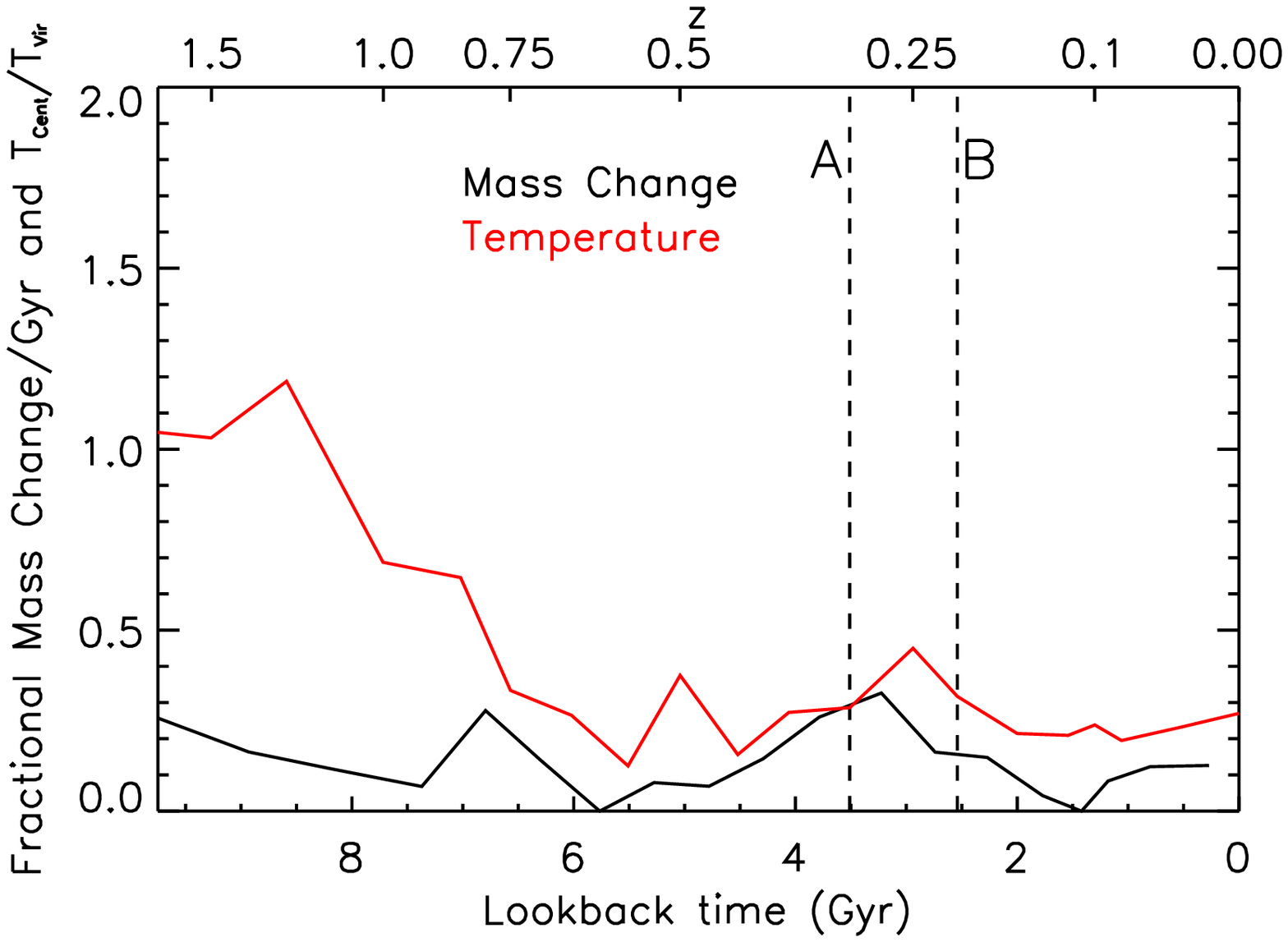}

\caption{Two snapshots of the history of a CC cluster with $M_{200} = 4.4 \times 10^{14} M_{\odot}$ and $T_{virial} =$ 3.7 keV at $z=0$.}
\label{CCevol}
\vspace{-8mm}
\end{center}  
\end{figure*}

In Figures \ref{NCCevol} and \ref{CCevol}, we show examples of the evolution of NCC and CC clusters, respectively, which well represent the general scenarios for how these clusters form.  Our simulations indicate that
lower mass clusters with $T<2$ keV form cool cores early in their
history when initial conditions produce central densities and
temperatures that allow the gas to radiatively cool.  This suggests
that many (most) lower mass clusters should have cool cores which is
consistent with the data in Figure \ref{fcc} in the next section and the
observations of poor clusters composed of early-type galaxies \cite[see
e.g., review by][]{Mul04,zabmul,ponman03,chen07}.  According to our
simulations, early mergers cause the fates of NCC and CC clusters to
diverge.

As shown by the representative example in Figure \ref{NCCevol}, NCC
clusters often undergo major mergers early in their history.  This
cluster had begun to develop a cool core at $z=1.25$ (see panel A).
However, the cluster experienced a major merger (mass increased by
$\approx$100\%) at $z \approx 1$ and the cool core was greatly
diminished (panel B).  Smaller mass halos with cool cores continue to
be accreted by this cluster but these CC's are ram pressure
stripped/disrupted usually within a single core passage (see panels C
and D).  By $z \approx 0.65$, there is no evidence of a CC associated
with the cluster dark matter density peak at this or later times.
Early mergers destroy the cool cores in NCC clusters, leaving behind
hotter, thermalized, moderately dense cores where the cooling time is
everywhere above the Hubble time. As shown in the next Section and in
the Appendix, the NCC cluster gas has become mostly relaxed within the
gravitational potential well (with minor perturbations from small
infalling halos) with a surface brightness profile well represented by
a $\beta$-model. Subsequently, cool halos infalling into these NCC
clusters do not survive passage through the central parts of the
clusters nor do the central conditions allow cool cores to
re-establish.  NCC clusters continue to experience minor mergers as
they now slowly evolve (typically, mass increases only $\approx$10\%
over Gyr time frames after $z \ge 0.5$ from multiple mergers for the
NCC as shown in Figure \ref{evol}). We suggest that such an early
major merger produced the characteristics observed today for the NCC
Coma cluster whose complex properties may be the result of previous
mergers \citep{burnscoma}.  

On the other hand, Figure \ref{CCevol} suggests that CC clusters evolve
differently.  This CC cluster had no significant change in mass until
$z=0.75$ and its only major merger did not occur until $z=0.3$.
Figure \ref{CCevol} shows the temperature and central cool core as the
merger is progressing (at $z=0.3$) (panel A).  The next snapshot at
$z=0.2$ shows the CC somewhat diminished but still present.  The
central temperature moved slightly upward but quickly readjusted
downward as the CC easily survives the shock heating and ram pressure
from the merger.  In contrast to the simple cooling flow model, CC
clusters may be no closer to hydrostatic equilibrium than NCC clusters
with the equivalent mass (see also Section 5.5).  This hierarchical
formation model for CC clusters makes clear predictions of
substructure and average cluster characteristics beyond the core that
are testable with X-ray data.  

\begin{figure}
\begin{center}
\includegraphics[width=3.5in]{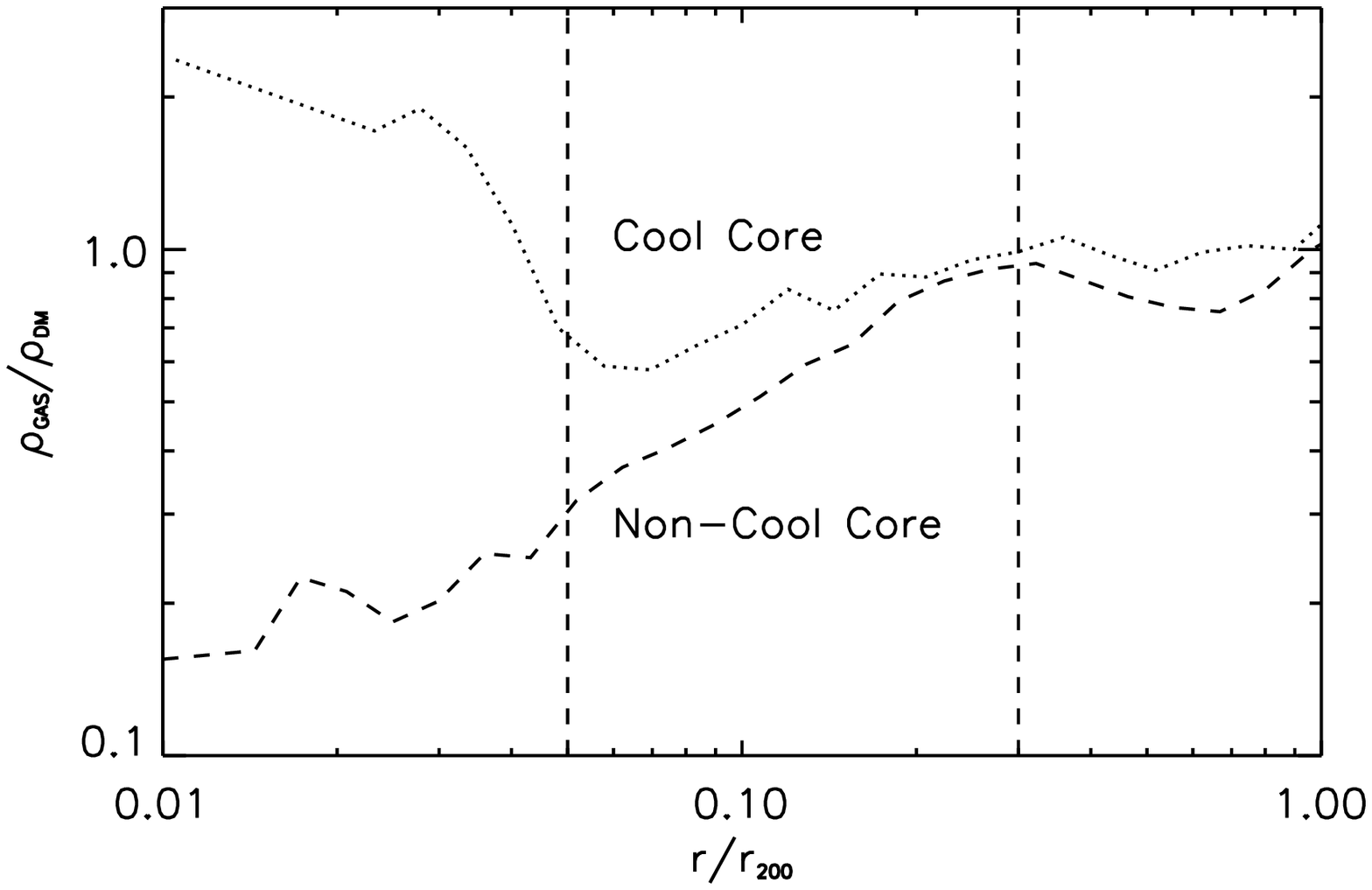}
\includegraphics[width=3.5in]{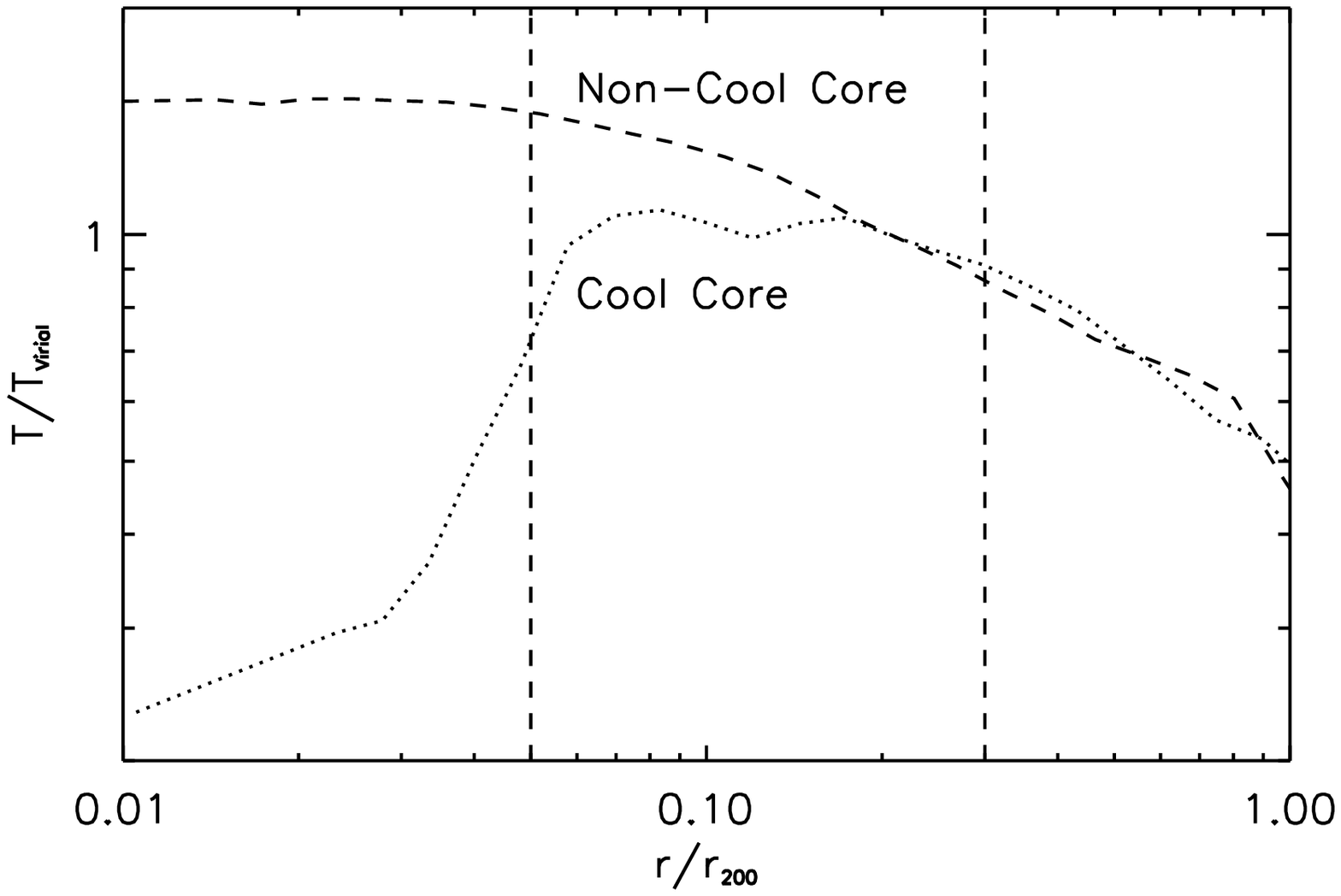}
\caption{Plots of the gas fraction and temperature profiles for
  representative CC and NCC numerical clusters.  The vertical lines
  designate the location of the newly defined CC cluster transition
  region ($0.05 r_{200}$ to $0.3 r_{200}$).  This transition region
  contains an excess of baryons compared to a typical NCC cluster and
  a relatively flat temperature profile, indicating different gas
  properties than either the core or the outer region.} 
\label{transition}
\vspace{-8mm}
\end{center}  
\end{figure}

Figure \ref{transition} shows the radial profiles of the baryon
fraction and the temperature for representative examples of numerical
CC and NCC clusters in our sample.  Even outside of the cool core
($\approx 0.05 r_{200} \approx 100 h^{-1}$ kpc, the first vertical
dotted line), there is an excess of baryons relative to NCC clusters
out to $\approx 0.3 r_{200}$ (second vertical dotted line).  (We note
that the dark matter density profiles are comparable for the CC and
NCC clusters.)  Such an extended ``transition region'' could be
created, in part, by gas ``sloshing" in the cluster potential well
following repeated mergers as proposed by \cite{shocks}. 

Figure \ref{transition} also shows the temperature differences between
each cluster type.  The NCC cluster demonstrates the universal
temperature profile that we described in \cite{loken02}.  The
temperature profile of the CC cluster rises steeply to $\approx 0.05
r_{200}$ and then it has a prolonged stretch of near-constant
temperature, again within the region $\approx 0.05 r_{200}$ to
$\approx 0.3 r_{200}$.  This CC cluster profile is less compact than that found for recent SPH simulations \citep{valdarnini} and agrees well with
observations \cite[see \eg][]{relaxed06, baldi}. 

Using these two trends, we define three components to a cool core
cluster: cool core, ``transition region'', and outer region.  The
transition region is differentiated by the excess of baryons outside
the core and relatively flat temperature profile (and low entropy)
compared to NCC clusters.  We have (subjectively) chosen the limits
$\approx 0.05 r_{200}$ to $\approx 0.3 r_{200}$ for this transition
region; $\approx 0.05 r_{200}$ is the traditional edge of the cool
core where the slope of temperature changes dramatically, and $\approx
0.3 r_{200}$ is an average location where the baryon fraction of each
type of cluster converges and the temperature begins to decrease.
Most current cluster X-ray observations also measure $S_X$ accurately
out to only $\approx 0.3 r_{200}$, meaning that most observations
measure predominantly the transition region in CC clusters (as will be
discussed further in \cite{gantner07}). 

These simulations predict a very different set of cluster
characteristics from those expected in the simple, non-evolving
cooling flow model or from cooling-only simulations.  Since CC and NCC
clusters have experienced different magnitudes and epochs of mergers,
there should be observational signatures remaining from the mergers.
In particular, we will show in the next section that the fraction of
clusters with cool cores is expected to be strong function of the
virial mass, in good agreement with recent observations.  We then will
show that single $\beta$-models systematically overestimate (or bias)
the densities and masses beyond the cores in CC clusters. Furthermore, our
simulations predict that more cool gas should be found beyond the
cores in CC clusters in comparison to NCC clusters.  Finally, this
scenario forecasts that CC rich clusters should be found in denser
supercluster environments at the present epoch.

\section{Consequences of Evolutionary Differences in CC and NCC
  Clusters} 

In this Section, we explore the differences in the
properties of CC and NCC clusters based upon the results of our
numerical simulations.  The simulations predict substantial
differences in the characteristics of these clusters beyond the cores.
These predictions can be tested with data from current and planned
X-ray telescopes. 

\subsection{Masses and Fractions of CC Clusters}

\begin{figure}
%\vspace{-10mm}
\begin{center}
%\epsscale{0.5}
\includegraphics[width=3.5in]{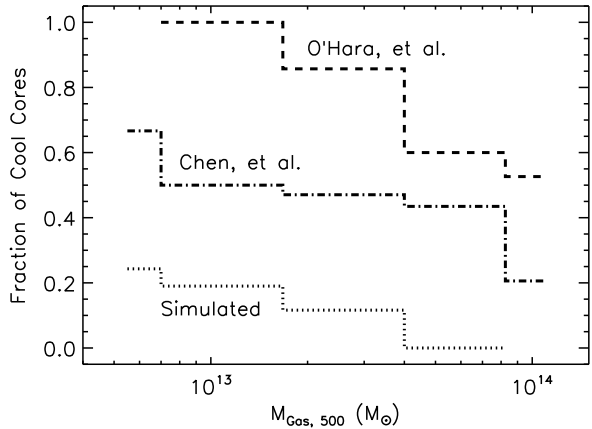}
%\vspace{-8mm}

\caption{Fraction of all galaxy clusters with cool cores as a function
  of mass.  The simulated data is from 1522 numerical clusters.
  \cite{chen07} is based upon their table of 106 observed clusters.
  \cite{ohara} is from their table of 45 observed clusters.  $M_{gas,
    500}$ was chosen as the measurement common across all three data
  sets. The differences in absolute fraction of CC clusters between
  the samples may have a variety of causes (\eg luminosity boosting by
  cool cores in flux-limited observational samples; numerical clusters
  may be affected by resolution effects, selection of cosmological
  parameters, or feedback prescription).} 
\label{fcc}
\end{center}  

\end{figure}

An intriguing new result is that the fraction of clusters with cool
cores is a strong function of cluster gas mass as shown in Figure
\ref{fcc}.  We display gas masses here instead of total cluster masses
to allow a direct comparison with the observation-derived data
presented by \citet{ohara} and \citet{chen07}. About a quarter of
simulated clusters with $M_{gas} \approx 5 \times 10^{12} M_{\odot}$
have cool cores whereas no high mass numerical clusters ($M_{gas} > 4
\times 10^{13} M_{\odot}$) have cool cores.  As a corollary to this
result, we find that the mean total mass for the 10 most massive CC
clusters at $z=0$ is $2.4 \pm 1.4 \times 10^{14} M_{\odot}$ whereas
the mean total mass for the 10 most massive NCC clusters is $11.3 \pm
4.0 \times 10^{14} M_{\odot}$.  These results support the idea that
cool cores are destroyed via multiple major mergers and the
probability of cool core disruption increases as clusters grow to the
size of the Coma cluster.  

A similar result can be seen from observational samples of clusters
using {\it ROSAT} data compiled by \citet{ohara} and by
\citet{chen07}. Both samples have a somewhat common ancestry from the
work of \citet{edge} with the samples consisting of nearby
($0.01<z<0.1$), moderate X-ray luminosity clusters.  In Figure
\ref{fcc}, we have plotted data from \citet{ohara} and \citet{chen07}
overlaid onto those for our sample of numerical clusters.  Although
the absolute values of the fractions differ between the observed and
numerical samples as discussed in Section 2, the general trend of
decreasing fraction of CC clusters with mass is present for both
observations and simulations. 

This finding is contrary to the expectations of the simple
non-evolving cooling flow model where the number of cool cores should
increase with cluster mass (as the central gas density
increases). Although observational selection effects are a possible
concern in the \citet{ohara} and \citet{chen07} catalogs, we believe
that this newly discovered trend contains important insights into the
formation of CC versus NCC clusters.   

\subsection{Evolution in the Fraction of Cool Cores?}

Recently, \citet{vikh06} reported that the number of observed CC
clusters declines dramatically to 15\% with redshift beyond $z \approx 0.5$ (versus 65\% for their nearby cluster comparison sample).  Because of the limited spatial resolution with {\it Chandra} at these distances, they use the central slope or ``cuspiness of the surface brightness'' to distinguish between CC and NCC clusters.

\begin{figure}
%\vspace{-10mm}
\begin{center}
%\epsscale{0.5}
\includegraphics[width=3.5in]{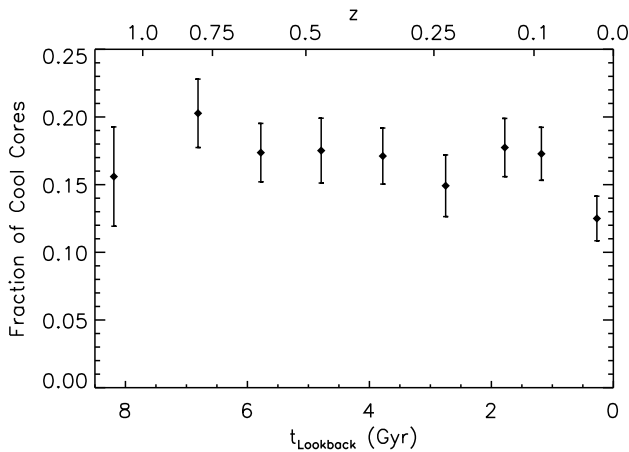}

\caption{Fraction of cool cores as a function of redshift for
  numerical clusters.  In contrast to \citet{vikh06}, we find no
  change in the fraction of cool cores with redshift.} 

\label{CCvsz}
\vspace{-8mm}
\end{center}  

\end{figure}

In Figure \ref{CCvsz}, we show the fraction within the co-moving volume of our numerical cool core clusters as a function of redshift for all clusters with $M > 10^{14} \mathrm{M_\odot}$ out to $z \approx 1$.  The error bars in each bin reflect the $\sqrt N$ uncertainties due to the number counts.  Within these errors, the fraction of CC clusters is not a strong function of redshift out to
$z \approx 1$ (15-20\%).  For $z > 1$, the fraction drops to $\approx$10\% but the dispersion is large because there are only a few CC clusters with $M > 10^{14} \mathrm{M_\odot}$ at these early epochs.

The flat distribution of CC fraction within the range $0 < z < 1$ is not inconsistent with the evolutionary formation scenario described in Section 4 and shown in Figure \ref{evol} for several reasons.  First, Figure \ref{evol} reveals that the greatest disparity in mass change between CC and NCC clusters occurs for $z>0.75$.  That is, most of the growth in NCC clusters via mergers occurs at the expense of CC clusters at earlier epochs.  For $z<0.75$, CC and NCC clusters grow at comparable rates.  Second, although CC clusters continue to be lost via mergers for $z < 0.75$, this is counterbalanced by the fact that the numbers of CC clusters above the mass cutoff of $10^{14} \mathrm{M_\odot}$ continue to increase because CC clusters also grow via accretion.  Thus, the rate at which CC clusters are destroyed is approximately equal to the rate at which new clusters are added above our $10^{14} \mathrm{M_\odot}$ mass limit.  This produces the effect of no apparent evolution in the fraction of CC clusters for $z<1$ in a mass-limited sample.

In order to explain the \citet{vikh06} result within
the context of our simulations, their sample would have to have
substantial selection effects possibly driven by the unique choice of
cool cores based upon the slope of the X-ray surface brightness
profile and resolution effects.  Alternatively, some form of
time-dependent baryon physics (\eg higher feedback rates at earlier
epochs as recently described by \cite{eastman07} for AGNs) not
incorporated into the present simulations could potentially boost the
fraction of CC clusters seen in the local Universe compared to earlier
epochs.

\subsection{Surface Brightness Profiles}

Turning next to the large-scale X-ray surface brightness profiles
($S_X$), we fit $\beta$-models in two different ways for two different
subsamples of the numerical clusters to examine potential differences
in the shapes and core radii of CC versus NCC clusters.  

\begin{figure}
\begin{center}
\includegraphics[width=3.5in]{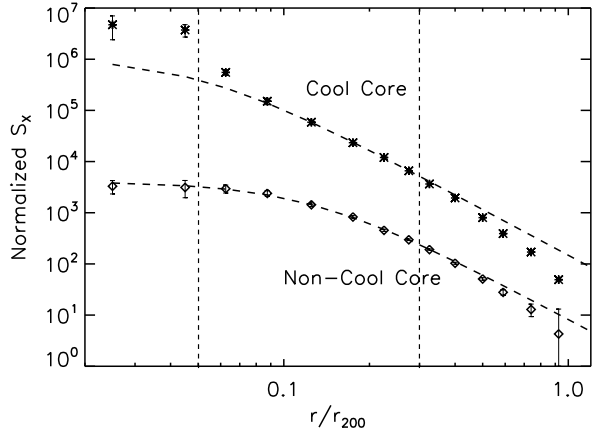}

\caption{Average synthetic X-ray surface brightness profiles for CC
  and NCC clusters with $z=0$ and $M > 10^{14} \mathrm{M_\odot}$ (10
  CC averaged together and 78 NCC averaged together).  Error bars are
  errors on the mean determined from variations within each bin.
  Dashed curves are the best fit $\beta$-models within the
  ``transition region'' (\ie between the vertical lines from $0.05
  r_{200}$ to $0.3 r_{200}$) for the CC average cluster profile and
  the best fit including all the points out to the right-most vertical
  line for the NCC cluster profile.  The NCC profile was arbitrarily
  shifted downward by a factor of 10 to better distinguish it from the
  CC profile.} 
 
\label{SX_profile}
\vspace{-6mm}
\end{center}  

\end{figure}

We began by producing average $S_X$ profiles from projected X-ray
images along a single axis for all the clusters at $z=0$ and with $M >
10^{14} \mathrm{M_\odot}$, separated into CC (10 clusters) and NCC (78
clusters) categories (several $z=0$ clusters were not used because of
contamination by multiple clusters).  The flux for individual profiles
was first normalized by ${M_{200}}^{3/2}$ (from the Mass-Temperature
scaling relationship as in \cite{ finoguenov01}) before averaging them
together.  We then fit a $\beta$-model to each of these average
profiles as would be done for observations.  That is, we fit the
profiles out to $0.3 r_{200}$ (roughly corresponding to $0.5 r_{500}
\approx 0.5$ Mpc) which is the typical limit to the observed surface
brightness in most X-ray exposures with current instruments (as we
discuss further in \cite{gantner07}).  For the average CC profile, we
excluded the cool core in making the $\beta$-model fit. The result is
shown in Figure \ref{SX_profile}.   

This figure indicates that the profiles (beyond the cool core) for CC
clusters are distinguished from NCC clusters in several important
ways.  First, the parameters for the $\beta$-models are different.
For the average CC profile, $r_c = 0.05 \pm 0.09$ $r_{200}$ and
$\beta=0.66 \pm 0.12$, whereas for the NCC average $S_X$ profile, $r_c
= 0.12 \pm 0.02$ $r_{200}$ and $\beta=0.66 \pm 0.07$.  That is, the
cluster core radii are much smaller for CC clusters, as also shown in
Figure \ref{statistics}.  There is also considerably more scatter in
the fit for the CC average profile (consistent with more variation
between individual profiles) than for the NCC $S_X$ profile.  Second,
the shape of the two $S_X$ profiles are different within the
transition region where the slope of the NCC is generally flatter than
the CC cluster, as would be expected from Figure \ref{transition}.  There are similar slope differences in the $S_X$ profiles between the NCC (Abell 401) and CC (Abell 85) clusters computed from deep Chandra observations reported by \cite{relaxed06}.  Third, the $\beta$-model is a better fit to the average NCC cluster profile than to the CC profile.  In particular, the $\beta$-model fit
to the regions that would be typically observed by current satellites
(\ie the transition region) for CC clusters significantly overshoots
the actual flux in the outer parts of clusters.  At $r_{200}$, the
$\beta$-model overestimates the flux of the average CC profile by a
factor of 3.8.  This will result in a serious bias of cluster gas
masses as we discussed in \cite{hall06}. 

\begin{figure}
\begin{center}
\includegraphics[width=3.5in]{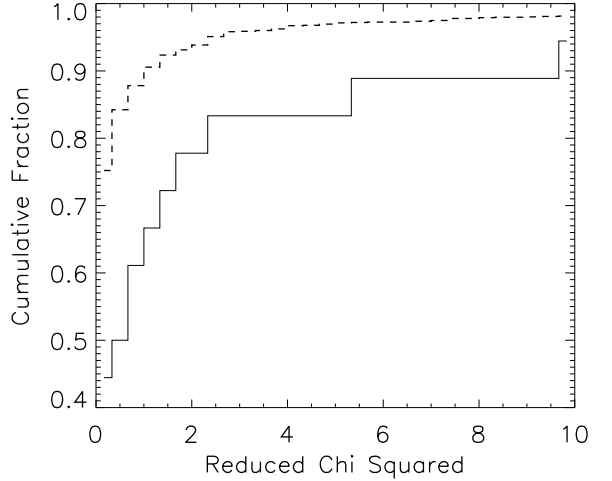}

\caption{Histograms of the reduced $\chi^2$ values for $\beta$-model
  fits for CC (solid) and NCC (dashed) clusters.  Each numerical
  cluster ($M_{200} > 5 \times 10^{14} M_{\odot}$, $z < 2$) is fit to
  $r_{500}$ and extrapolated out to $r_{200}$, and a $\chi^2$
  goodness-of-fit is calculated for the entire profile. } 

\label{chisq_hist}
\vspace{-8mm}
\end{center}  

\end{figure}

We also fit $\beta$-models to individual profiles for all numerical CC
and NCC clusters (from a single projection) in our master database
with $M > 5 \times 10^{14} \mathrm{M_\odot}$ and $z < 2$.  In this
case, we fit models out to $r = r_{500}$.  We did not use the inner
portion of the profiles dominated by the cool core (determined by the
point where the slope of the temperature profile becomes negative) in
making fits to CC clusters.   We then calculated the reduced $\chi^2$
goodness-of-fit values (including extrapolations of the fits out to
$r_{200}$) as compared with the numerical X-ray profiles.  A histogram
of those reduced $\chi^2$ values for CC and NCC clusters is shown in
Figure \ref{chisq_hist}.  As also indicated in Figure
\ref{SX_profile}, the $\beta$-models fit the NCC clusters much better
than the CC clusters.  About 88\% of the NCC clusters have $\chi^2
<1$, whereas about one-third of the CC clusters have $\chi^2 >1$.
Once again, this is caused by the slope changes from the transition
region to the outer core in the CC clusters which is not fit well by a
single $\beta$-model. 

As we show in the Appendix, good $\beta$-model fits to $S_X$ suggest
that a  nonisothermal gas in the intracluster medium (ICM) is in
approximate (but not necessarily perfect, see Section 5.5) equilibrium
with the gravitational potential well of the cluster (under the assumption that the ICM gas is polytropic).  So, the above
results suggest that the ICM in NCC clusters is approximated by a gas
with a balance between heating and cooling that is quasi-relaxed in an
NFW-like dark matter potential.  This is consistent with Figure
\ref{evol} which shows that NCC clusters underwent major mergers early
in their history but have subsequently settled into a
quasi-equilibrium state with only minor on-going mergers.  This
contrasts with CC clusters that have temperature and density profiles
inconsistent with a simple adiabatic gas well beyond the cool core. 

\citet{relaxed06} recently analyzed the gas and total mass profiles
for 13 nearby, ``relaxed'' clusters with temperatures between 0.7 and
9 keV using data from {\it Chandra}.  All of the clusters in this
sample have cool cores.  They examined the surface brightness profiles
out to at least $r_{500}$ and concluded that the profiles are ``not
described well by a beta model''.  When single 
${\beta -}$models are fit to the inner portions of the clusters (but
excluding the cool cores), they poorly extrapolate the gas density and
mass profiles in the outer parts of the CC clusters as we found for
the numerical clusters.  In addition, \citet{relaxed06} observed that
the cooler regions in low temperature clusters are confined to a
smaller fraction of the virial radius than in the hotter CC clusters.
This is consistent with trends found in our numerical simulations as
described by \citet{hall06} and by \citet{akah}. Finally, Vikhlinin et
al. note that low temperature (mass) clusters ($T <$ 2.5 keV) with
cool cores have a bigger ratio of central to virial temperature than
do the clusters with larger $T_{virial}$.  We plan to explore the
origin of this effect with new higher resolution simulations with more
sophisticated heating prescriptions.

\subsection{Temperatures and Hardness Ratios Beyond the Cluster Cores}

\begin{figure}
\begin{center}
\includegraphics[width=4.5in]{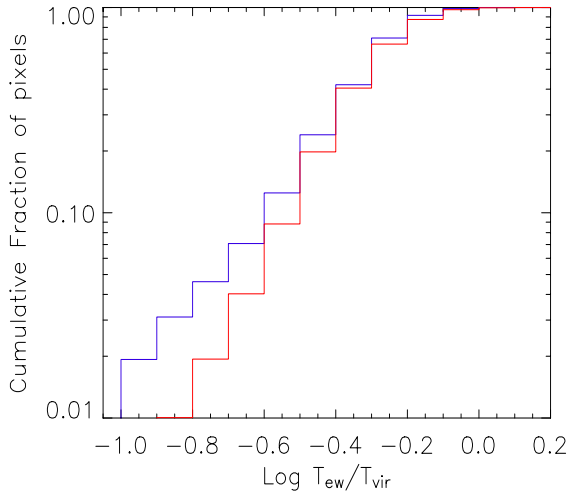}

\caption{Histograms of the ratio of emission-weighted temperatures
  ($T_{ew}$) from $r = 180 h^{-1}$ kpc to $r_{200}$ divided by
  $T_{virial}$ from temperature images for CC and NCC clusters.
  Clusters from the numerical catalog are included with $M_{200} > 5
  \times 10^{14} M_{\odot}$.  Red are NCC and blue are CC clusters.} 

\label{T_hist}
\vspace{-3mm}  
\end{center}  

\end{figure}

The temperature profiles for the two clusters shown in Figure
\ref{transition} indicate that the temperature distributions for CC
and NCC clusters are substantially different out to $\approx 0.3
r_{200}$.  In particular, the broad ``transition region'' has both a
cooler and flatter distribution of temperature outside the cool core
for the CC cluster in comparison to the NCC cluster.  How general is
this result for the total sample of numerical clusters?  To address
this question, we made emission-weighted temperature images of the
clusters in our numerical cluster catalog with $M_{200} > 5 \times
10^{14} M_{\odot}$ and $0<z<0.5$.  From these images, we produced the
histogram of temperatures (normalized by $T_{virial}$), excluding cool
cores, shown in Figure \ref{T_hist}.  The distribution of temperatures
beyond the cool cores for CC clusters is significantly different from
NCC clusters with a broad tail toward lower temperatures.  CC clusters
have $\sim$40\% more gas with $T_{ew} < 0.3 T_{virial}$ beyond the
cores than NCC clusters. 

\begin{figure*}
\begin{center}
\includegraphics[width=5.0in]{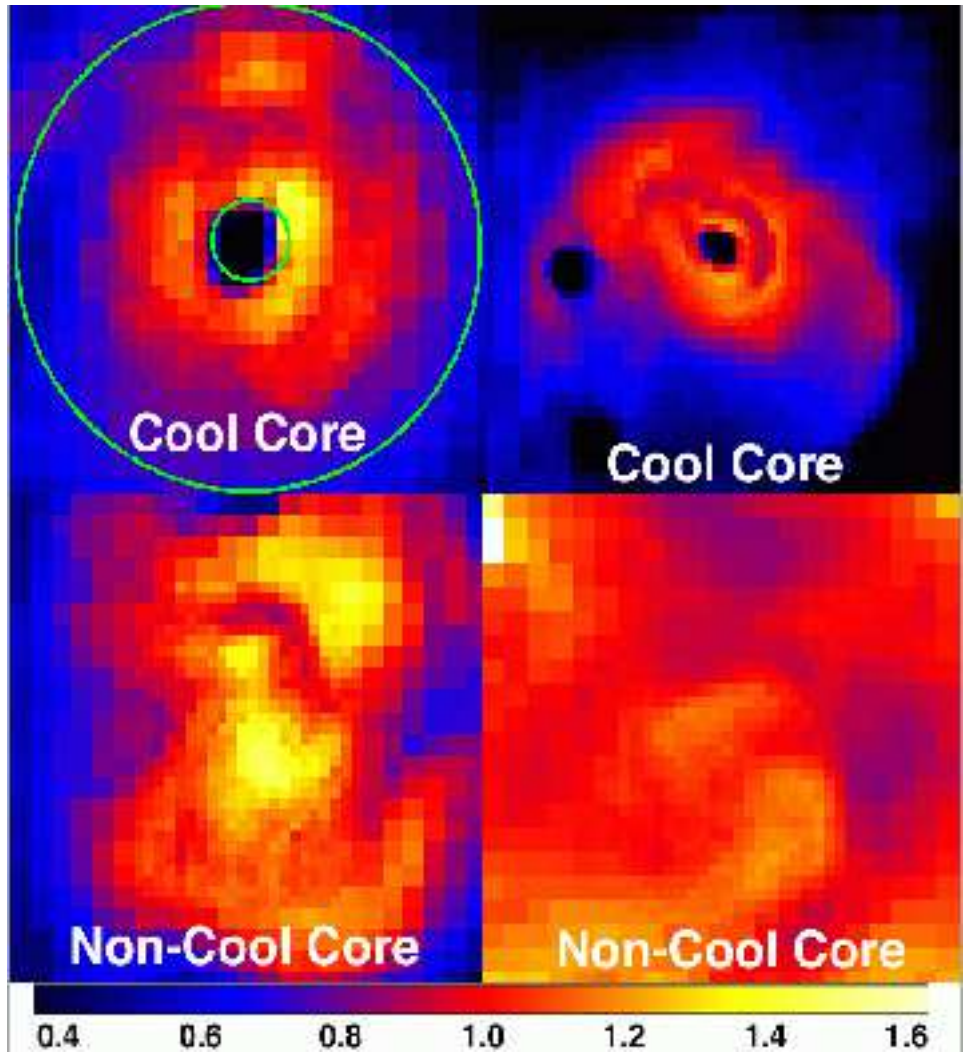}

\caption{A sample of X-ray hardness ratio (HR) maps from 4 different
  simulated clusters (2 CC on top and 2 NCC on bottom) with  fields of
  view of $0.33 r_{200}$.  The two circles in the upper left image
  delineate the ``transition region'' between $0.05 r_{200}$ to $0.3
  r_{200}$ as described in Figure \ref{transition}.  Each HR map was
  made by dividing an image of the X-ray surface brightness from 2 to
  8 keV by an image from 0.5 to 2.0 keV (i.e., typical {\it Chandra}
  hard and soft bands).  Each is then normalized by the hardness ratio
  corresponding to the cluster virial temperature so that values $<1$
  represent gas with $T < T_{virial}$. Note that non-central cool
  ``blobs'' are generally infalling halos. } 

\label{HRmaps}
\end{center}  
\vspace{-8mm}  

\end{figure*}

\begin{figure}
\begin{center}
\includegraphics[width=3.5in]{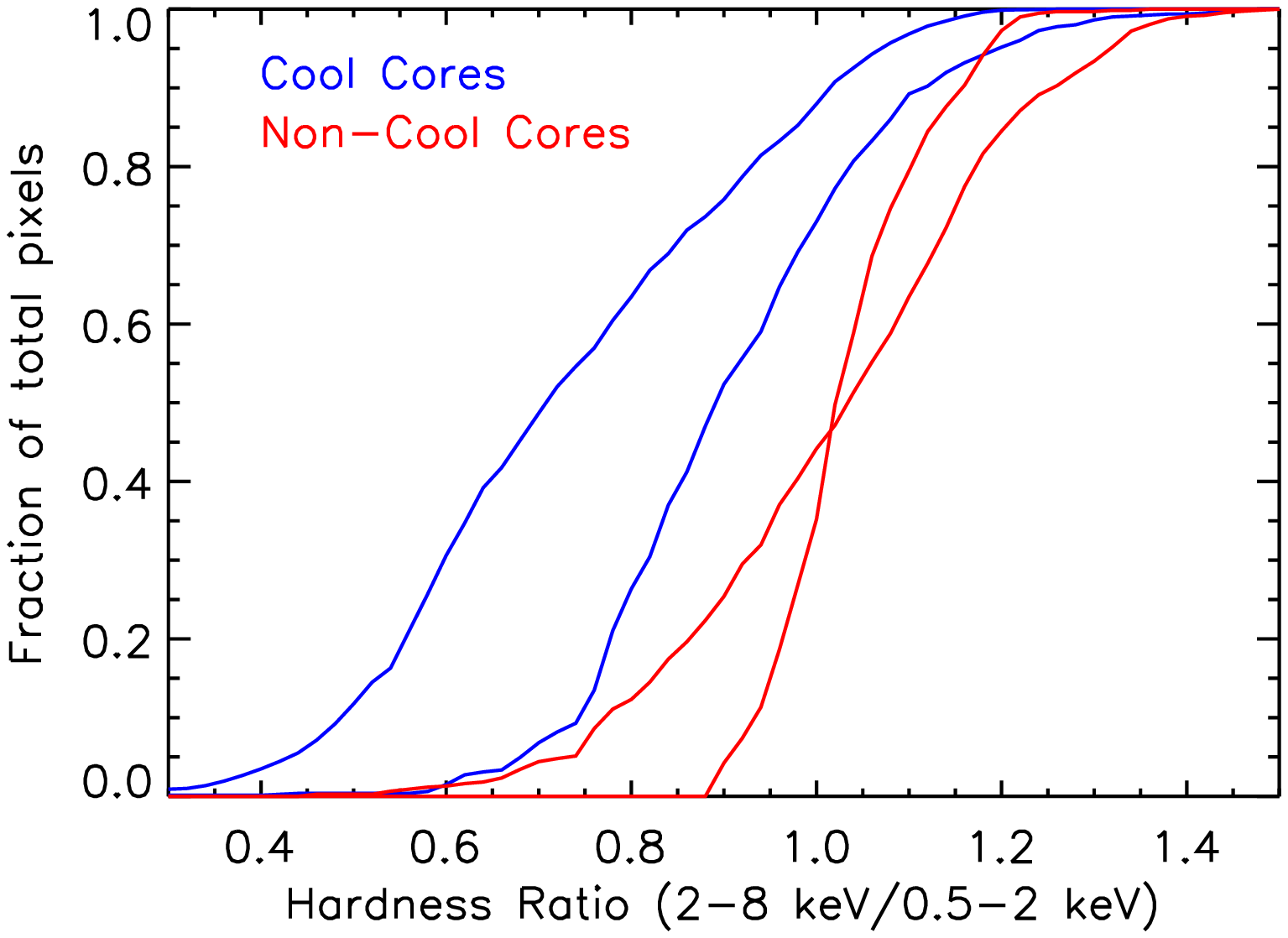}

\caption{The cumulative fraction of pixels below each hardness ratio
  value for the sample of simulated clusters in Figure
  \ref{HRmaps}. The plot was assembled for regions between $0.05-0.3
  r_{200}$ (excludes cool core in CC clusters).  The NCC clusters are
  centered roughly at 1, meaning that most of their gas is
  approximately equal to $T_{virial}$.  The CC clusters are noticeably
  cooler in this transition region.} 

\label{HRplots}
\end{center}  

\end{figure}

We predict that this signature will be apparent in hardness ratio maps
that are commonly made from X-ray observations.  As shown in Figure
\ref{HRmaps} for four typical cases drawn randomly from our
simulations, the hard-to-soft band ratios (2-8 keV/0.5-2 keV) do a
good job of illustrating the abundance of cooler gas beyond the cores
in CC clusters.  Figure \ref{HRplots} shows the cumulative fraction of
pixels below a given hardness ratio for all four clusters in Figure
\ref{HRmaps}.  For the two CC clusters, we have excised the cool cores
($<0.05 r_{200}$) so as not to bias the results with gas already known
to be cooler than its NCC counterpart.  As expected, the two CC
clusters have a majority of pixels  with values $<1$ and therefore are
cooler in the transition region than the NCC clusters.  The NCC
clusters are both centered approximately at hardness ratio $\approx
1$, hence the gas in these clusters is roughly at the virial
temperature of the clusters. 

We shall show in \cite{gantner07} that there is very good agreement in
the predicted hardness ratios from our simulations with X-ray
observations of clusters from the {\it Chandra} and {\it ROSAT}
archives.

\subsection{The Supercluster Environments of CC and NCC Clusters}

The mass and temperature evolution plots in Figure \ref{evol} indicate
that NCC clusters underwent major mergers early in their history in
contrast to the milder accretion over time for CC clusters.  This may
also suggest that the larger scale environments in which these two
types of clusters live are different since accretion of halos and
diffuse material must come from the cosmic web.  It is possible that
NCC clusters began their lives in higher overdensity regions which
then accelerated the growth of these clusters via mergers \cite[see
\eg][]{mowhite96, gao05}. 

In an effort to explore the possible influence of the differences in
the supercluster environments for CC and NCC clusters, we calculated
the real space densities (which are expected to correlate with
accretion rates) of all halos with $M_{200} > 10^{13} M_{\odot}$
($\approx$ mass resolution of simulations) within a radius of $5
r_{200}$ of numerical rich clusters with virial masses $1-6 \times
10^{14} M_{\odot}$ (approximate mass range of the CC clusters as shown
in Figure \ref{masstemp}).  We calculated these densities for a series
of redshifts between 0 and 1.5, and separated clusters by CC and NCC
according to our definitions in Section 2. 

\begin{figure}
\begin{center}
\includegraphics[width=3.5in]{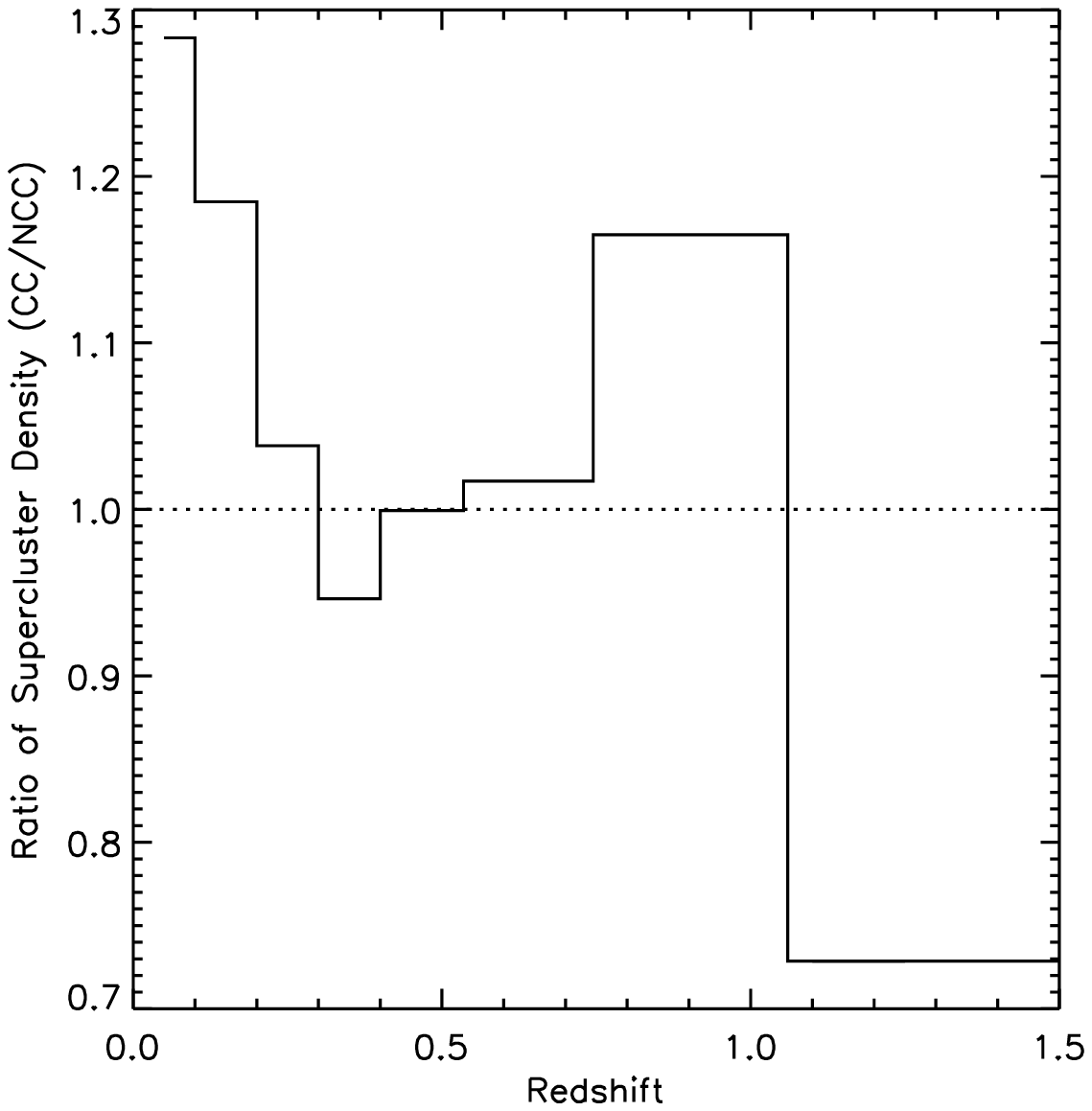}

\caption{Histograms of the ratio of the real space densities of halos
  surrounding numerical clusters with $M_{200} = 1-6 \times 10^{14}
  M_{\odot}$ for CC versus NCC clusters.} 

\label{super_ratio}
\vspace{-8mm}
\end{center}  

\end{figure}

The ratio of supercluster densities for CC to NCC clusters as a
function of redshift is shown in Figure \ref{super_ratio}.  At early
epochs ($z>1$), the average supercluster density is somewhat higher
around NCC clusters than CC clusters.  One might expect clusters that
are experiencing major bouts of accretion of subclusters that result
in the destruction of embryonic cool cores to be surrounded by a
higher density of halos.  At times corresponding to $0.7 < z < 1$,
there may a slight underdensity of halos around NCC clusters in
comparison to CC clusters as one might expect if the NCC clusters
suffered a large amount of mass accretion effectively ``clearing out''
its nearby neighborhood.   Interestingly, at late times ($z < 0.3$),
the density of halos around CC clusters is $\approx$30\% greater than
for NCC clusters.  However, unlike earlier epochs where the mass ratio
of the main cluster to the average neighboring cluster is often
$\approx$ a few, this average mass ratio of the rich cluster to the
halos at $z < 0.3$ is much larger as the main cluster has grown
considerably over the past 10 Gyr.  This means that there are many
small subclusters falling into the CC clusters but their relative
impact is small compared to that for NCC clusters at earlier epochs
(see Figure \ref{transition}).  However, the above trends are weak at
best and there is a large dispersion in supercluster densities between
individual clusters. 

It is interesting to note that this possible trend of overabundance of
halos around numerical CC clusters at the present epoch is also found
in Abell clusters.  \citet{loken99} constructed a volume-limited
sample of $z < 0.1$ Abell clusters that was estimated to be 98\%
complete in an effort to investigate their supercluster environs.
They separated clusters into CC and NCC.  They found that CC Abell
clusters have twice the density of neighboring clusters as do NCC
clusters out to a radius of 43 $h^{-1}$ Mpc.  We attempted to mimic
the \citet{loken99} analysis by recomputing halo densities out to
radii of 43 $h^{-1}$ Mpc and included neighbor halos in the
calculation only if they had masses $> 10^{14} M_{\odot}$ (\ie
Abell-like clusters).  We found that the density of neighboring halos
is $\approx$40\% greater for CC than NCC clusters, slightly larger
than in Figure \ref{super_ratio}, but still considerably less than
what \citet{loken99} propose for Abell clusters. 

\subsection{Deviations from Hydrostatic Equilibrium}

Galaxy clusters are potentially powerful tools for precision
cosmology.  Accurate cluster mass determinations and gas fractions,
along with cluster abundance counts, can provide key constraints on
the dark energy parameter, $w$, as well as $\Omega_b$, $\Omega_m$, and
$\sigma_8$ \cite[\eg][]{wangstein, haiman01}.  In recent efforts,
samples of CC clusters are being used exclusively  because they are
believed to be dynamically relaxed.  For example, \cite{allen07}
selected a sample of 42 hot, X-ray luminous clusters with $0.05 < z
<1.1$, all of which have short central cooling times ($<$ a few $10^9$
yrs), to constrain cosmological parameters from $f_{gas}$.  Previous simulations \cite[\eg][and references therein]{rasia06, nagai06} have called into question hydrostatic equilibrium for clusters.  But, are CC clusters really more dynamically relaxed than NCC clusters?  Figures \ref{evol} and \ref{super_ratio} seem to call this assumption
into question. 

\begin{figure}
\begin{center}
\includegraphics[width=3.5in]{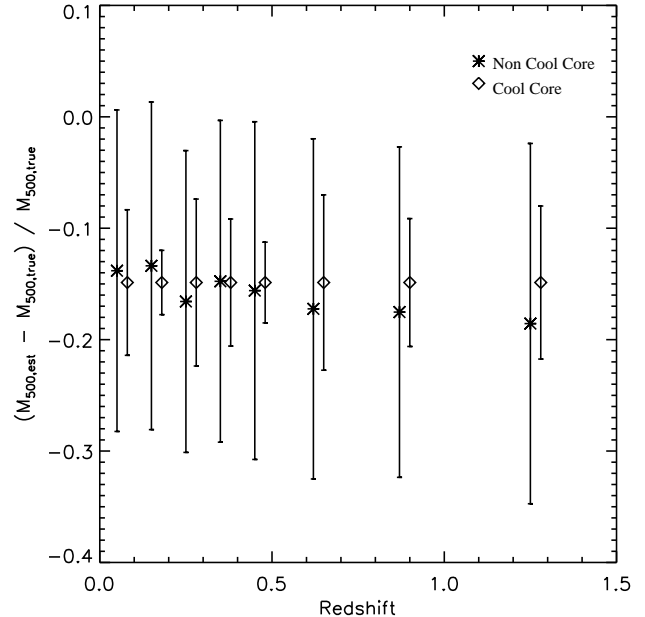}

\caption{The fractional deviation of cluster masses estimated assuming
  hydrostatic equilibrium versus the true virial masses as a function
  of redshift.  The standard deviations in the distributions are also shown.} 

\label{hse_z}
\vspace{-8mm}
\end{center}  

\end{figure}

To explore this further, we calculated the deviations from hydrostatic equilibrium for all the clusters in our numerical archive with $M_{200} > 10^{14} M_{\odot}$ \cite[see also][]{jeltema07}.  We did this by calculating the estimated mass of clusters from the gradients in the temperatures and gas densities in the usual way assuming hydrostatic equilibrium.  In this case, we have used the spherically averaged profiles of temperature and density from the three-dimensional simulated cluster data. We have therefore eliminated any systematic effect resulting from conversion of observed quantities. So we expect that the
resulting ``hydrostatic masses'' should be closer to the true values
than ones which would be observationally derived. We then compared
these hydrostatic masses to the true mass for each cluster.  We did
this for a series of redshift intervals between 0.0 and 1.5, and
separated clusters between CC and NCC.  The result is shown in Figure
\ref{hse_z}. 

The average estimated cluster masses assuming hydrostatic equilibrium
are biased low for all the clusters by $\approx$15\%.  This bias is
constant for CC clusters at different redshifts but appears to be
slightly worse for NCC clusters at earlier epochs ($\approx$19\%).  In
addition to the bias, the scatter in these mass estimates is high.
20\% to 30\% underestimates are possible at the 1$\sigma$ level.
Importantly, and what is new here, CC clusters are no better than NCC clusters as biased mass indicators.  Both are equally low, although the scatter in CC
clusters is about half that of the NCC clusters. 

\cite{shocks} similarly cast doubt on hydrostatic equilibrium in CC
clusters due to the common presence of observed cold fronts and the
inferred ``gas sloshing''.  For the clusters in our simulations, we
find that the kinetic energy of bulk gas motions contributes at the
$\approx$10\% level compared to the total energy \cite[see
also][]{rasia06}. 

If our numerical clusters are representative of real clusters, the
apparent significant deviations from hydrostatic equilibrium for both
CC and NCC clusters must be considered in choosing to use them for
precision cosmology estimators.

\section{Summary and Conclusions}

Galaxy clusters are complicated, generally non-equilibrium systems
where nongravitational physics is important in the cores.  To the best
of our knowledge, no previous numerical simulations have been able to
produce both cool core (CC) and non-cool core (CC) clusters in the
same numerical volume.  Our heating and cooling prescription (with an
approximate balance between heating and cooling), however, has
resulted in a simulation with both CC and NCC clusters.  The
temperature profiles of our numerical CC clusters qualitatively match
observations, although the central gas densities are higher than
observed.  Our fraction of cool cores is low compared to that of
recent observed samples.  On the other hand, the distinction in
$\beta$-model parameters ($r_c$ and $\beta$) between CC and NCC
clusters observed in samples of real clusters is reflected in our
numerical clusters.  Similarly, the distributions of gas fraction and
gas mass with emission-weighted projected temperatures agree fairly
well with observations.  Overall, our numerical clusters have general
characteristics that concur with X-ray data of observed clusters. 

We propose an answer to the question posed in the title of this paper,
\ie only some clusters have cool cores because of evolutionary
differences driven by early major mergers.  Our numerical simulations
suggest that the histories of cool core and non-cool core clusters are
significantly different.  Our NCC numerical clusters suffer early
major mergers when nascent cool cores are destroyed.  CC clusters, on
the other hand, grow more slowly without early major mergers.  CC
clusters have a broad ``transition region'' in their gas distribution
extending between the cool core and a radius of $\approx 0.3 r_{200}$
where the gas fraction is higher than for NCC clusters and the
temperature profile is nearly isothermal.  This transition region and
difference in evolution lead to a number of testable predictions for
X-ray observations of real clusters. 

We find that the fraction of cool core clusters is a strong function
of mass with fewer CC clusters at higher gas masses.  This general
trend qualitatively agrees with the analysis of recent nearby X-ray
cluster samples by \cite{ohara} and \cite{chen07}.  On the other hand,
we do not find any significant variation in the fraction of numerical
cool cores with redshift in contrast to the recent claim by
\cite{vikh06}. 

The X-ray surface brightness profiles for NCC clusters are well fit by
single $\beta$-models whereas the outer emission for CC clusters is
biased low compared to $\beta$-models.  The resulting gas densities
and gas masses of CC clusters estimated from single $\beta$-model
extrapolations are biased high by factors of 3-4. 

CC clusters have $\approx$40\% more cool gas beyond the cores within
the transition region than do NCC clusters.  This results in a very
different distribution of X-ray hardness ratios beyond the cool core
for CC versus NCC clusters.  We predict that such differences will be
observable with current X-ray imagers. 

There are some indications that the supercluster environs for CC and
NCC clusters are different from each other today and in past epochs.
At $z>1$, NCC clusters appear to have more halos in their
neighborhoods than CC clusters.  At $z < 0.3$, this trend is reversed
with more halos around CC clusters.  This separation between CC and
NCC clusters for low z clusters qualitatively agrees with supercluster
density calculations for nearby Abell clusters. 

Finally, we find that both CC and NCC clusters are biased low in their
mass estimation by $\approx$15\% assuming hydrostatic equilibrium.  In
this sense, it appears that CC clusters are no better than NCC
clusters as mass estimators, unlike what is generally assumed.  This
is important to consider in using CC clusters for precision
estimations of cosmological parameters. 

In an upcoming paper \citep{gantner07}, we will compare the above
predictions with X-ray observations of rich clusters from both the
{\it Chandra} and {\it ROSAT} archives.  The initial agreement is
quite good. 
 
There are some important remaining issues with the current
simulations.  We  plan to address these with a series of new numerical
simulations at higher resolution to overcome the limitations of the
current computational set of clusters.  We will explore the impact of
$\Omega_b$ and $\sigma_8$ on the fraction of cool cores that are
produced in the computational volume.  We will also refine our
heating/cooling prescription to better match current observational
constraints.  Once the influence of these factors is understood on the
creation of CC and NCC clusters, the fraction of cool core clusters
could serve as an important new constraint on cluster baryonic physics
and/or dark energy models. 

\vspace{5mm}
This work was supported in part by grants from the National Science
Foundation (AST-0407368) and the NASA ADP (NNX07AH53G) program.  We
thank Brian O'Shea, T. Reiprich, M. Voit, and M. Markevitch for
stimulating discussions.  We acknowledge the referee for providing useful comments and suggestions.  We also appreciate the Aspen Center for
Physics for hosting several of the authors (JOB, EJH) where some of the final
work on this project was completed.

\appendix
\section{Appendix: X-ray Profile for an Adiabatic ICM in an NFW Cluster Potential}

What does a good fit to the X-ray surface brightness profile by a
$\beta$-model imply about the dynamical state of the cluster gas when
it is nonisothermal? When the intracluster gas is relaxed in a
\cite{NFW97} (NFW) dark matter potential (derived from N-body
simulations) and in hydrostatic equilibrium, we can solve a simple
equation for its radial distribution  \cite[see
also][]{NFW97,makino99}. The equation for hydrostatic equilibrium is a
simplification of the Euler equations for an ideal fluid, setting
fluid velocity to zero gives 
\begin{equation}
\nabla P_{gas} = -\rho_{gas} g, 
\end{equation}
where $P_{gas}$ is the pressure of the ICM gas, $\rho_{gas}$ indicates
the gas density, and $g$ is the local gravitational
acceleration. Under the assumption of spherical symmetry, we can
simplify this to
\begin{equation}
\frac{dP_{gas}}{dr} = -\rho_{gas}(r) g(r). 
\end{equation}
We assume here that the dark matter potential dominates
the gravitation, and do not include the contribution of the gas to the
potential, which should result in only minor error.  In that case, we
can write $g(r)$ from an NFW dark matter profile as
\begin{equation}
g(r) = \frac{G M_{<R}}{r^{2}},
\end{equation}
where $G$ represents the universal gravitational constant, and
$M_{<R}$ indicates the dark matter mass inside the radius of
interest. That mass can be calculated by integrating the NFW profile
\begin{equation}
\rho_{dm}(x) = \rho_{0,dm} \frac{1}{x(1+x)^2}
\end{equation}
where
\begin{equation}
x = \frac{r}{r_c}
\end{equation}
and $r_c$ is the core radius and $\rho_{0,dm}$ is the central
normalization of the profile. Integrating the profile to get the total
enclosed mass 
\begin{equation}
M_{<R} = 4\pi {r_c}^3 \rho_{0,dm} \int \frac{1}{x(1+x)^2} x^2 dx
\end{equation} 
We assume that the gas follows a nonisothermal, adiabatic equation of state such that
\begin{equation}
P = k \rho^{\gamma}.
\end{equation}
where k is a constant.
Then, the hydrostatic equilibrium equation to be solved can be written as
\begin{equation}
\frac{dP_{gas}}{dr} = -P^{1/\gamma}  \frac{4\pi G {r_c}^3
  \rho_{0,dm}}{r^2}  \left[\int \frac{1}{x(1+x)^2} x^2 dx \right].
\end{equation}

\begin{figure}
\begin{center}
\includegraphics[width=0.6\textwidth]{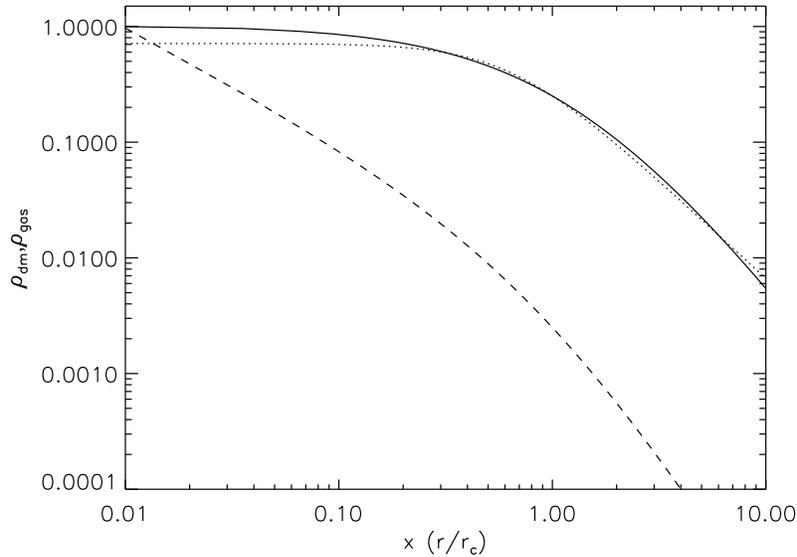}
\caption{Density profiles described in the text. Dashed is the
  standard NFW dark matter density profile in arbitrary units, solid
  line is the resulting gas density profile from integrating the
  equation of hydrostatic equilibrium in the NFW potential. The dotted
  line is a standard $\beta$-model fit to the gas density profile.}
\label{nfw}
\vspace{-8mm}
\end{center}
\end{figure}

Numerical integration of Eq. 8 results in the profiles shown in Figure \ref{nfw}. The
solid line is the solution to Eq. 8, the dashed is the NFW dark matter
density profile, and the dotted is a standard $\beta$-model fit to the
gas density.  Note that the $\beta$-model fits Eq. 8 very well for $r
> 0.2 r_c$ but the $\beta$-profile is somewhat flatter in slope within
the core in comparison to Eq. 8.  This suggests that good
$\beta$-model fits to X-ray profiles imply gas that is relatively
relaxed within the dark matter potential.

%===============================================================================
% References
%===============================================================================
%\input{cc.bbl}

%===============================================================================

\end{document}